\documentclass[a4paper,11pt]{article}
\pdfoutput=1

\usepackage{jcappub,xcolor} 

\usepackage{graphicx}
\usepackage{epstopdf}
\usepackage{epsf}
\usepackage{bm}
\usepackage{amsmath,mathtools,commath}
\usepackage{amsfonts}
\usepackage{amssymb}%

\usepackage{xcolor}
\usepackage{hyperref}

\providecommand{\U}[1]{\protect\rule{.1in}{.1in}}

\newcommand{\be}{\begin{equation}}
\newcommand{\ee}{\end{equation}}

\newcommand{\mincir}{\raise
-3.truept\hbox{\rlap{\hbox{$\sim$}}\raise4.truept\hbox{$<$}\ }}
\newcommand{\magcir}{\raise
-3.truept\hbox{\rlap{\hbox{$\sim$}}\raise4.truept\hbox{$>$}\ }}

\newcommand{\ud}{\,\mathrm{d}}
\newcommand{\uD}{\,\text{D}}
\usepackage{comment}

\newcommand{\calO}{{\cal O}}

\newcommand{\calR}{{\cal R}}

\newcommand{\beq}{\, \begin{equation} }
\newcommand{\eeq}{\, \end{equation}}

\begin{document}

\hfill{\small APCTP-Pre2023-010}

\title{Enhanced power spectra from multi-field inflation}

\author[a]{Perseas Christodoulidis,}
\affiliation[a]{Department of Science Education, Ewha Womans University, Seoul 03760, Korea}

\author[a,b]{Jinn-Ouk Gong}
\affiliation[b]{Asia Pacific Center for Theoretical Physics, Pohang 37673, Korea}
\emailAdd{perseas@ewha.ac.kr; jgong@ewha.ac.kr}

\abstract{
We investigate the enhancement of the power spectra large enough to produce primordial black holes in models with multiple scalar fields. 
We present analytic solutions for the perturbations in the case of constant turns without the need for an effective field theory for the first time and clarify the role of the Hubble friction that has been overlooked previously.
We derive the criteria for an arbitrary number of fields that can lead to an exponential amplification of the curvature perturbation on subhorizon scales, while leaving the perturbations stable on superhorizon scales. Finally, we apply our results to a three-field generalization of the ``ultra-light'' scenario and show how the presence of field-space torsion can yield distinct observables compared to the two-field case. 
}

\maketitle

\tableofcontents

\section{Introduction}


The physical evolution of the observed universe is very well described by the hot big bang universe. Supported by solid observational evidences including the expansion of the universe, abundance of light elements, and the cosmic microwave background (CMB), the concordant cosmological model -- called $\Lambda$CDM -- specified by six parameters exhibits surprising agreement with the most recent observations on the CMB~\cite{Planck:2018vyg}. Yet, a successful hot big bang evolution demands a set of extremely finely tuned initial conditions. One representative example of such conditions is the homogeneity and isotropy of the CMB beyond the causal patches. These difficulties regarding the initial conditions can be resolved by introducing, before commencing the standard hot big bang, a period of accelerated expansion of the universe, called inflation \cite{Starobinsky:1980te,Sato:1981ds,Sato:1980yn,Kazanas:1980tx,Guth:1980zm,Linde:1981mu,Albrecht:1982wi}. During inflation the expansion is so fast such that the whole observed universe was once in a single causally connected patch, which explains the homogeneity and isotropy of the CMB. Furthermore, quantum fluctuations during inflation are stretched and become the primordial perturbations, which seed the observable structure in the current universe. The properties of these primordial perturbations have been constrained by cosmological observation programmes with increasing accuracy during the last decades, and are consistent with most recent data~\cite{Planck:2018jri}.


While the general paradigm of inflation is robust, the realization of inflation is not an easily surmountable problem. To implement a period of inflation, we usually invoke a scalar field, called the inflaton, with a sufficiently flat potential energy that dominates the energy density of the universe due to the Hubble friction. Then the inflaton ``slowly rolls'' down the potential and behaves effectively as a cosmological constant, leading to the exponential expansion of the universe -- inflation~\cite{Mukhanov:2005sc,Weinberg:2008zzc,Lyth:2009zz}. Under this slow-roll approximation, the calculations of the two-point correlation functions of the primordial perturbations, or the power spectra, are straightforward and we can easily constrain the functional form of the inflaton potential by data from the anisotropies of the CMB and the large-scale inhomogeneous distribution of galaxies~\cite{Lidsey:1995np}.

However, in the standard model of particle physics (SM) that has been verified and confirmed experimentally, there is no fundamental scalar field that can play the role of the inflaton.\footnote{Note that by introducing a non-minimal coupling to gravity, the SM Higgs field may play the role of the inflaton~\cite{Bezrukov:2007ep}.} Thus, the realization of inflation resorts to the theories beyond SM, or in the context of effective field theory. One common feature of these theories is the existence of degrees of freedom with heavier mass scales than our interest. Since it is very likely that the energy scale of the early universe is far higher than that proved by terrestrial particle accelerators, as high as $10^{15}$ GeV, those heavy degrees of freedom which play little role in the SM could have actively participated in the inflationary dynamics. Then, these participating degrees of freedom are relevant for inflation and thus can all be identified as the inflatons. Therefore there are good motivations to consider multiple degrees of freedom during inflation -- multi-field inflation. For reviews see e.g.~\cite{Bassett:2005xm,Wands:2007bd,Langlois:2010xc,Gong:2016qmq}.


Furthermore, recent cosmological observations have brought tantalizing phenomenological hints indicating that we need more than the six-parameter standard cosmological model. There are several outliers in the power spectrum of the temperature fluctuations of the CMB, which could be accommodated by ``features'' in the inflationary dynamics caused by various reasons~\cite{Chluba:2015bqa}. The rotation angle of the CMB polarizations is found to be non-zero~\cite{Minami:2020odp,Eskilt:2022cff}, which calls for possible parity-violating mechanism coupled to the photons such as the Chern-Simons term with axion~\cite{Carroll:1989vb}. The value of the Hubble parameter determined by the local measurements including supernovae is not in agreement with the value determined by the CMB with very high significance~\cite{Riess:2019cxk}. Currently there is no leading candidate to resolve this discrepancy, but many possibilities have been suggested including e.g. early domination of dark energy~\cite{Poulin:2018cxd}. All these observations indicate that the standard cosmological model needs to be refined, including the constituents yet to be discovered.


Especially, the discovery of gravitational waves (GWs) by LIGO~\cite{LIGOScientific:2016aoc} has spurred many studies on various aspects of GWs and related topics. An exciting possibility is that the binary black holes relevant for the merger events are not the remnants of the stellar evolution but of primordial origin -- primordial black holes (PBHs)~\cite{Zeldovich:1967lct,Hawking:1971ei,Carr:1974nx}. It is believed that PBHs are formed when the amplitude of the primordial perturbations at certain length scale is so high that as soon as such a region with very high perturbation amplitude enters the horizon after inflation, gravitational collapse immediately follows and the corresponding horizon-size black hole is formed. Therefore, for the PBHs, we need a large enhancement of the power spectrum compared to the value constrained on the CMB scales, $\calO(10^{-5})$. The exact number necessary for the generation of PBHs depends on the details of the numerical studies, but generally it is estimated that the value should be $\calO(10^{-1} - 10^{-2})$~\cite{Harada:2013epa}. Thus, we need an amplification of the primordial perturbations by $\calO(10^3 - 10^4)$  around the scale that corresponds to the size, or equivalently the mass, of the PBHs.

Such a large enhancement in single-field inflation is, however, not arbitrary but is limited with the steepest growth of the power spectrum being $k^4$~\cite{Byrnes:2018txb,Carrilho:2019oqg}. The resulting PBH mass and the prospects for LIGO and future observations are highly constrained. In multi-field inflation, the amplification of the primordial perturbations has also been discussed in many different contexts, e.g. oscillations in the inflationary trajectory~\cite{Pi:2017gih}, hybrid-inspired models \cite{Garcia-Bellido:1996mdl,Cheong:2019vzl,Braglia:2020eai,Braglia:2022phb,Cheong:2022gfc}, (sudden) turns in field space \cite{Palma:2020ejf,Fumagalli:2020nvq,Fumagalli:2020adf,Kolb:2022eyn,Anguelova:2020nzl,Bhattacharya:2022fze,Fumagalli:2021mpc,Aragam:2023adu}, resonance in the speed of sound~\cite{Cai:2018tuh}, multiple features~\cite{Boutivas:2022qtl} to list only a few examples.

In this article, we investigate the general conditions for the enhancement of the power spectra of perturbations during multi-field inflation.
This article is outlined as follows. In section~\ref{sec:multi_general}, we first provide a concise review of multi-field inflation to introduce our notations and necessary materials for the following discussions. In section~\ref{sec:analytical_sol} we derive  analytic solution for the enhancement of the curvature perturbation in two-field models that undergo constant turns. In section~\ref{sec:hamilton} we describe the conditions under which perturbations for an arbitrary number of fields may experience a temporary exponential enhancement in the subhorizon regime but remain under control on superhorizon scales. In section~\ref{sec:isometries} we investigate the models with isometries and focus on ultra-light isocurvature perturbations. Finally, in section~\ref{sec:conclusions} we summarize our study. 

\section{Multi-field perturbations}  
\label{sec:multi_general}

The purpose of this section, devoted to a short and self-contained review, is four-fold. First, we introduce the notations we will adopt throughout this article. Second, we set up the Lagrangian to be analyzed in the next sections. Third, we discuss some ideas relevant to the following contents, and lastly we adopt the Hamiltonian formulation of inflationary perturbations.

\subsection{Orthonormal frame and initial conditions} 
\label{subsec:orthonormal}

Our starting point is a $\mathcal{N}$-dimensional multi-field system with a generic field space metric $G_{IJ}$ coupled to the Einstein gravity:
\begin{equation}
S = \int d^4x \sqrt{-g} \bigg[ \frac{R}{2} 
- \frac{1}{2}G_{IJ} g^{\mu\nu} \partial_\mu \phi^I \partial_\nu \phi^J - V(\phi) \bigg]
\, .
\end{equation}
The second-order action for the gauge-invariant perturbations 
\begin{equation}
Q^I \equiv \delta \phi^I + {\psi \over H} \dot{\phi}^I \, ,
\end{equation}
is given by the following expression~\cite{Gong:2011uw}:
\begin{equation} 
\label{eq:quadr_act}
S_2 = { 1 \over 2} \int \ud t \ud^3x a^3 \left( \uD_t Q^I \uD_t Q_I -{1 \over a^2 } \nabla Q_I  \nabla Q^I  - M_{IJ}  Q^IQ^J \right) \, ,
\end{equation}
where the field index is raised and lowered by the field space metric $G_{IJ}$, and the mass matrix $M_{IJ} $ is defined as follows~\cite{Achucarro:2010da}:
\begin{align} 
\label{eq:mass_matrix}
M_{IJ} & \equiv V_{;IJ} - R_{IKLJ}\dot{\phi}^K \dot{\phi}^L + (3-\epsilon) \dot{\phi}_I \dot{\phi}_J + {1 \over H} \left(  \dot{\phi}_I V_{,J}  +   \dot{\phi}_J V_{,I}  \right) \, ,
\end{align}
with a semicolon being a covariant derivative with respect to $G_{IJ}$ and $\epsilon \equiv -\dot{H}/H^2$. Here, $\uD_t$ is the total derivative operator acting on vectors as
\begin{equation}
\uD_t A^I \equiv {\ud \over \ud t} A^I + \Gamma^I_{JK} \dot{\phi}^K A^J \, ,
\end{equation}
and for the field space Riemann tensor we adopt the following convention:
\begin{equation}
R^I_{KLJ} \equiv \partial_{L} \Gamma^I_{KJ} - \partial_{J} \Gamma^I_{KL} + \Gamma^I_{LN}  \Gamma^N_{KJ} - \Gamma^I_{JN}  \Gamma^N_{KL}  \, . 
\end{equation}
Recall that the field fluctuations behave as vectors in the field space~\cite{Gong:2011uw} and the use of the covariant derivative operator $\uD_t$ ensures that the action is covariant under arbitrary field redefinitions.

For computational ease we change to a local orthonormal frame~~\cite{Achucarro:2010da}. This can be done by defining a new set of orthonormal basis vectors $\{ \hat{\boldsymbol{e}}_a \}$ according to 
\begin{equation} 
\label{eq:orthonormal1}
\begin{split}
\hat{\boldsymbol{e}}_a & \equiv E^{~I}_{a} \boldsymbol{e}_I \, ,
\\
\boldsymbol{e}_I  & \equiv \tilde{E}^{~a}_{I} \hat{\boldsymbol{e}}_a \, ,
\end{split}
\end{equation}
where $\tilde{E}^{~a}_{I}$ is the inverse matrix of $E^{~I}_{a}$, i.e.~$\tilde{E}^{~a}_{J}  E^{~I}_{a} = \delta^{~I}_{J}$ and $E^{~I}_{a} \tilde{E}^{~b}_{I} = \delta^{~b}_{a}$, with the following properties:
\begin{equation}
\begin{split}
E^{~I}_{a} E_{b}^{~J} G_{IJ} & = \delta_{ab} \, , 
\\
\tilde{E}^{~a}_{I}  \tilde{E}^{~b}_{J}  \delta_{ab} & = G_{IJ}  \, , 
\\
G_{IJ} E^{~J}_{a} & = \delta_{ab} \tilde{E}^{~b}_{I} \, .
\end{split}
\end{equation}
The last equation relates the components of $\tilde{E}$ with the metric tensor and $E$. This new set of basis vectors constitutes a non-coordinate basis, as the commutator between those basis vectors does not vanish: $[\hat{\boldsymbol{e}}_a,\hat{\boldsymbol{e}}_b]\neq 0$.  
The next step is to define a covariant derivative in the new frame. Taking the covariant time derivative of \eqref{eq:orthonormal1} gives
\begin{equation} 
\uD_t \hat{\boldsymbol{e}}_a = 
 \left( \dot{E}^{~I}_{a} +  E^{~J}_{a} \Gamma^I_{JK} \dot{\phi}^K \right) \tilde{E}^{~b}_{I} \hat{\boldsymbol{e}}_b  \, ,
\end{equation}
and this suggests the definition of a new matrix $z_{a}{}^b$ as
\begin{equation} 
z_{a}{}^b  \equiv  \left( \dot{E}^{~I}_{a} + E^{~J}_{a}  \Gamma^I_{JK} \dot{\phi}^K \right) \tilde{E}^{~b}_{I} \, ,
\end{equation}
which is antisymmetric in its two indices\footnote{Note that for tensors defined in the orthonormal base, we raise and lower indices using the Euclidean metric $\delta_{ab}$ so there is no distinction between the two indices. However, we keep track of the position of indices in each column.} and can be used to define a connection with respect to the non-coordinate basis:
\begin{equation}
\uD_t \hat{\boldsymbol{e}}_a = z^{~b}_{a} \hat{\boldsymbol{e}}_b \, .
\end{equation} 
Until now, our discussions on the change to a set of orthonormal basis is completely general. From now on, we identify the orthonormal basis as the kinematic one so that the first two basis vectors are tangent and normal to the background trajectory~\cite{Gordon:2000hv,Achucarro:2012sm}.
The antisymmetric matrix $z$ has non-zero components only above and lower of the principal diagonal, which allows us to write the previous system of equations as
\begin{equation}
\uD_{t} 
\begin{pmatrix}
\hat{\boldsymbol{e}}_{\rm t} \\
\hat{\boldsymbol{e}}_{\rm n} \\
\hat{\boldsymbol{e}}_{\rm b} \\
\vdots 
\end{pmatrix} 
=
\begin{pmatrix}
0 & \omega & 0   &\cdots \\
-\omega & 0 & \tau_{\rm b}  & \cdots \\
0 &  -\tau_{\rm b} &0 &   \cdots \\
\vdots  &\vdots   & \vdots  & \ddots  
\end{pmatrix} 
\begin{pmatrix}
\hat{\boldsymbol{e}}_{\rm t} \\
\hat{\boldsymbol{e}}_{\rm n} \\
\hat{\boldsymbol{e}}_{\rm b} \\

\vdots  
\end{pmatrix} 
\, ,
\label{eq:FrenetSerretEfolds}
\end{equation}
also known as the Frenet-Serret system \cite{kreyszig2013differential}. The first three basis vectors are known as the tangent, normal and binormal vectors and the parameters $\omega$ and $\tau_{\rm b}$ are the curvature (commonly known as the turn rate in the inflationary literature)  and the torsion respectively. In the orthonormal basis the components of perturbations are expressed as 
\begin{equation}
Q^I  = E^{~I}_{a} q^a \, , 
\end{equation}
so that the covariant derivative of $q^a$ is given by
\begin{equation}
\uD_t \pmb{q} = \uD_t (q^a  \hat{\boldsymbol{e}}_a) 
= \left( \dot{q}^a + z^{~a}_{b} q^b \right) \hat{\boldsymbol{e}}_a = \left( \dot{q}^a - z^a_{~b} q^b \right) \hat{\boldsymbol{e}}_a \, ,
\end{equation}
with second-order action: 
%
%
\begin{equation}  
\label{eq:quadratic_lagrangian}
S_2 = { 1 \over 2} \int \ud t \ud^3x a^3 
\left[ \delta_{ab}\dot{q}^a \dot{q}^b  - 2 z_{ab} \dot{q}^a q^b   - {1 \over a^2 } \delta_{ab} \nabla q^a  \nabla q^b  - \left( M_{IJ}E^{~I}_{a} E^{~J}_{b} + z_{ac} z^{c}_{~b} \right) q^a q^b \right] \, ,
\end{equation}
and the equations of motion in Fourier space:
\begin{equation} 
\ddot{q}^a + \left( 3 H \delta^a_{~b} - 2 z^a_{~b}\right)  \dot{q}^b   + \left( {k^2 \over a^2 } \delta^a_{~b}  + M_{IJ}E^I_{~a} E^J_{~b} + z^a_{~c} z^{c}_{~b}  -3 H z^a_{~b} - \dot{z}^a_{~b} \right) q^b =0 \, .
\end{equation}
%

By an appropriate field redefinition, and using the Frenet-Serret equations, we can simplify the Lagrangian \eqref{eq:quadratic_lagrangian}. By redefining the tangential component as $q_{\rm t} \equiv \sqrt{2\epsilon} \mathcal{R}$ and performing integrations by parts, the Lagrangian simplifies to 
%
\begin{equation}
\begin{aligned} 
L  
& =  
a^3 \left\{ {1 \over 2} \dot{q}^a \dot{q}_a  - {1 \over 2 a^2 } \nabla q^a  \nabla q_a  - {\epsilon  \over  a^2 }  \nabla \mathcal{R}  \nabla \mathcal{R}  -  z_{ab} \dot{q}^a q^b  +{1 \over 2} \omega^2 q_{\rm n}^2  + \epsilon \dot{\mathcal{R}}^2  - 2 \omega \sqrt{2\epsilon} \dot{\mathcal{R}} q_{\rm n}  
\right.
\\
& \quad \left.
- {1\over 2} \left(M_{ab} + z_{ac} z^{c}_{~b} \right)q^aq^b   + \left(M_{\mathcal{R}{\rm b}} + \omega \tau_{\rm b} \right)\sqrt{2\epsilon}\mathcal{R} q_{\rm b}  + \left(  M_{\mathcal{R}{\rm n}} + 3 H \omega + \dot{\omega} +{\dot{\epsilon} H \over \epsilon} \right) \sqrt{2\epsilon}\mathcal{R} q_{\rm n}
\right.
\\
& \quad \left.
-  \left[ M_{\mathcal{R}\mathcal{R}} - \omega^2  - {1\over 4} \eta^2 H^2 - {3 \over 2} H \dot{\epsilon} 
- {1 \over \epsilon} { \ud \over \ud t }\left( {\epsilon \eta H \over 2} \right) \right] \epsilon\mathcal{R}^2    \right\}  \, ,
\end{aligned}
\end{equation}
where now $a,b\neq \mathcal{R}$ and $\eta \equiv \dot\epsilon/(H\epsilon)$. Using the Frenet-Serret equations we find that the second and third lines which involve $M_{\mathcal{R}\mathcal{R}}$, $M_{\mathcal{R}\rm{n}}$ and $M_{\mathcal{R}\rm{b}}$ vanish identically. Thus, the quadratic Lagrangian for multi-field perturbations in terms of the curvature perturbation $\mathcal{R}$ and the isocurvature perturbations $q_i$ is 
\begin{equation}
\begin{aligned}
\label{eq:multi_lagrangian}
L  &=  a^3 \left[{1 \over 2} \dot{q}^i \dot{q}_i -  z_{i j} \dot{q}^i q^j - {1 \over 2 a^2 } \nabla q^i  \nabla q_i  - {1 \over 2}\left(M_{ij}  + z_{im} z^{m}{}_{j} \right) q^i q^j + {1\over 2}\omega^2 q_{\rm n}^2
\right.
\\
& \quad \left.
- 2 \omega \sqrt{2\epsilon} \dot{\mathcal{R}} q_{\rm n}   - {\epsilon  \over  a^2 }  \nabla \mathcal{R}  \nabla \mathcal{R}    + \epsilon \dot{\mathcal{R}}^2  \right] \, .
\end{aligned}
\end{equation}
Similar expressions for the second-order Lagrangian have been derived in \cite{Achucarro:2018ngj,Pinol:2020kvw}.

To properly assign initial conditions for the perturbations (e.g.~Bunch-Davies) we need to transform the action into its canonical form. This can be done by using a transformation to another local orthonormal basis which can eliminate the kinetic couplings between the fields. We will work with the conformal time $\ud \tau \equiv a^{-1}\ud t$ and redefine the fields as $u^i \equiv a q^i $ to simplify the action \eqref{eq:quadratic_lagrangian} to
\begin{equation} 
\label{eq:quadratic_conformal}
S_2 = { 1 \over 2} \int \ud \tau \ud^3x  \left[ \delta_{ij} {\ud u^i \over \ud \tau}{\ud u^j \over \ud \tau}  - 2 aH Z_{ij} {\ud u^i \over \ud \tau} u^j  -  \delta_{ij} \nabla u^i  \nabla u^j - (aH)^2 m_{ij}  u^i u^j \right] \, ,
\end{equation}
where we discarded total derivative terms and used for convenience the following definitions:
\begin{equation}
\begin{split}
Z_{ij} & \equiv \frac{z_{ij}}{H} \, ,
\\
\mathcal{M}_{ij} & \equiv \frac{M_{ij}}{H^2} \, ,
\\
m_{ij} & \equiv \mathcal{M}_{IJ} E_{i}^{~I} E_{j}^{~J}  + Z_{il} Z^{l}_{~j} -\left(2 - \epsilon \right) \delta_{ij} \, .
\end{split}
\end{equation}
%
The expression \eqref{eq:quadratic_conformal} is not yet suitable for quantization due to the second term that mixes the fields and their derivatives. We can eliminate these terms by introducing explicit time dependence in the definition of the orthonormal basis vectors \cite{Achucarro:2010da,Lalak:2007vi,Cremonini:2010ua}. A time-dependent change of basis
\begin{equation}
u^i = R^i_{~\alpha}(\tau) v^{\alpha} \, ,
\end{equation}
will eliminate the cross terms if $R$ satisfies the following differential equation:
\begin{equation}
\label{eq:orthogonal_matrix}
{\ud R^{i}_{~\alpha} \over \ud \tau} - aH Z^i_{~j} R^j_{~\alpha} = 0 \, ,
\end{equation}
i.e.~if the matrix is orthogonal. Regarding the initial condition of \eqref{eq:orthogonal_matrix}, we are completely free to choose any convenient, physically relevant time. In accordance with the previous literature we consider that the two basis vectors are aligned at horizon crossing $k \tau_* = -1 $ and hence the initial condition becomes $R^{i}_{~\alpha}(\tau_*) = \delta^i{}_{\alpha}$. In terms of the new fields we find a canonical second-order action as
\begin{equation} 
\label{eq:quadratic_conformal_2}
S_2 = { 1 \over 2} \int \ud \tau \ud^3x  \left[ \delta_{\alpha \beta} {\ud v^\alpha \over \ud \tau}{\ud v^\beta \over \ud \tau}  -  \delta_{\alpha \beta} \nabla v^\alpha  \nabla v^\beta + (aH)^2 (2 -\epsilon) \delta_{\alpha \beta } v^{\alpha}v^{\beta}  - (aH)^2 \mathcal{M}_{ij} R^i_{~\alpha}R^j_{~\beta} v^{\alpha}v^{\beta} \right] \, ,
\end{equation}
with the following equations of motion in Fourier space:
\begin{equation} \label{eq:subhorizon_v}
{\ud^2  \over \ud \tau^2} \boldsymbol{v} + \left[ k^2 - (aH)^2 (2 -\epsilon) \right] \boldsymbol{v} + (aH)^2 \boldsymbol{R}^T \boldsymbol{\mathcal{M}}\boldsymbol{R} \boldsymbol{v}  =0 \, .
\end{equation}
%
For these fields we can assign the usual positive energy initial condition, i.e. the Bunch-Davies mode function solution on subhorizon scales, $k/(aH) \approx - k \tau \gg 1$:
\begin{equation}
v^{\alpha}|_{\rm BD} = {1 \over \sqrt{2k}}e^{-ik\tau}\left(1 - {i \over k\tau} \right) \, ,
 \qquad \alpha = 1,\cdots,\mathcal{N} \, ,
\end{equation}
and find the initial conditions for the $u$ fields as
\begin{equation}
u^i|_{-k\tau\gg1} = R^i_{~\alpha}(\tau,\tau_*) v^{\alpha}|_{\rm BD} \, .
\end{equation}

\subsection{Scale dependence of the power spectrum}  
\label{subsec:scale_dependence}

At sufficiently late times we are interested in the scale dependence of the normalized power spectrum defined by
\begin{equation}
\left\langle {\mathcal{R}} (\boldsymbol{k}) {\mathcal{R}} (\boldsymbol{k}')  \right\rangle 
\equiv (2\pi)^3 \delta^{(3)} \left( \boldsymbol{k} - \boldsymbol{k}' \right) 
{ 2 \pi ^2 \over k^3 } \mathcal{P}_\mathcal{R} 
\, .
\end{equation}
Due to the complexity of the coupled equations, in general one has to solve numerically the coupled equations of perturbations to find the superhorizon value of $\mathcal{R}$. However, we will show that when every background quantity, i.e.~$\epsilon$, $\mathcal{M}_{\rm nn}\cdots$, is constant the scale dependence of the power spectrum can be read off the equations of perturbations without explicitly solving them. This can be done in the following way: First, in the differential equations for the independent variables of \eqref{eq:subhorizon_v} we trade $\tau$ with~\cite{Gong:2001he}
\begin{equation} 
\label{eq:xdefinition}
x \equiv - k \tau \, ,
\end{equation}
which for this type of solutions (i.e.~of scaling type) it becomes $x = (1-\epsilon)k/(aH)$, yielding 
\begin{equation} 
{\ud^2  \over \ud x^2} \boldsymbol{\tilde{v}} + \left[ 1- {2 -\epsilon \over (1 -\epsilon)x^2}  \right] \boldsymbol{\tilde{v}} + {1\over x^2 }\boldsymbol{R}^{-1} \boldsymbol{\mathcal{M}}\boldsymbol{R} \boldsymbol{\tilde{v}}  =0 \, .
\end{equation}
Notice that in this case the rotation matrix becomes a function of $x$, i.e.~$R=R(x)$, and so independent of $k$. 
Second, we rescale the fields as $\tilde{v}=\sqrt{k}v$ and in this way the initial conditions for each mode become identical, since they do not depend on the wavenumber. Moreover, the $\tilde{v}$ fields remain normalized as their past asymptotic solution for $x \rightarrow \infty$ is given by $\tilde{v}^a|_{x \rightarrow  \infty} = \exp(ix)/\sqrt{2}$, and they satisfy the Wronskian condition:
\begin{equation}
\tilde{v} \tilde{v}_{,x}^* - \tilde{v}^* \tilde{v}_{,x} = - i \, .
\end{equation}
With these two transformations we observe that each mode $k$ satisfies exactly the same differential equation with the same initial conditions. Therefore the superhorizon, asymptotic values $\tilde{v}_{\rm as}$ will be the same for all modes. From these we can calculate the asymptotic value of the curvature perturbation as
\begin{equation}
\label{eq:R_scale}
\mathcal{R}_{\rm as} =  {H \over \sqrt{2\epsilon k^3}} x R_{~\alpha}^1 \tilde{v}^\alpha_{\rm as} \, ,
\end{equation}
%
and the $k$-dependence of the power spectrum comes exclusively through $H$ when it is expressed as a function of $x$:
\begin{equation}
\mathcal{P}_\mathcal{R} \sim H^2(k,x)  \, .
\end{equation}
For scaling solutions $H$ evolves as $H = H_0 e^{-\epsilon N}$ and 
we can easily find the relation between $N$ and $x$ as
\begin{equation}
(1 - \epsilon) N = - \log x + \log \left( \frac{k}{a_0 H_0} \right) \, ,
\end{equation} 
with $a_0$ setting the point $N=0$. This yields the $k$-dependent $H$ of the form
\begin{equation}
H = H_0 x^{\epsilon /(1-\epsilon)} \left({k \over a_0 H_0 } \right)^{- \epsilon /(1-\epsilon)} \, ,
\end{equation}
%
and the normalized power spectrum scales as $ \mathcal{P}_\mathcal{R}  \sim k^{-2\epsilon/ (1-\epsilon) }$, \textit{irrespective of the number of fields}. This remarkably has exactly the same form as the single-field power spectrum for scaling solutions (see e.g.~\cite{Stewart:1993bc} for the calculation in the single-field case).

When $H$ and the slow-roll parameters $\epsilon$ and $\eta$ are time-dependent but slowly evolving, the differential equation will acquire $k$-dependence suppressed by the slow-roll parameters. This dependence is present in the components of the mass matrix as well as in the Hubble radius term which now reads $k/(aH) \approx -k\tau/(1+\epsilon)$  (see also  appendix~\ref{app:massive_field}) 
to lowest order in the slow-roll parameters. If we are interested in the leading-order computation of the power spectrum, we can safely ignore terms involving $\epsilon$ and $\eta$  as their contributions to $\mathcal{R}_{\rm as}$ will be logarithmic or higher order. The power spectrum becomes
\begin{equation}
\mathcal{P}_\mathcal{R} = {H^2 \over 4\pi^2} \frac{x^2 | \tilde{u}_{\rm t} |^2}{\epsilon} \, .
\end{equation}
Assuming that the isocurvature perturbations have decayed on superhorizon scales we can write
\begin{equation}
| \tilde{u}_{\rm t} |^2 \approx \tilde{u}_{i}^* \tilde{u}_{i} = \tilde{v}_{\alpha}^* \tilde{v}_{\alpha} \, ,
\end{equation}
and the final result does not depend on the rotation matrix $R$. If $\epsilon$ and $\eta$ vary slowly, but all other quantities are constant then the scale dependence of the power spectrum becomes identical to the single-field case to lowest order.\footnote{This has been noticed before in cases where the orthogonal perturbation is heavy and can be integrated out, leading to an effective description of the curvature perturbation with a modified speed of sound \cite{Achucarro:2010da}. However, our conclusion is not limited to heavy orthogonal fields.}
Otherwise, 
to find the asymptotic value of the curvature perturbation one has to solve the differential equations explicitly, which is the topic of section \ref{subsec:broadturns}.



\subsection{Hamiltonian description}

From now on, we adopt the Hamiltonian description of perturbations, rather than scrutinizing the equations of motion for perturbations. Because the action of our interest is quadratic, the resulting equations of motion consist a linear system. Thus, in the Hamiltonian formulation solving for the perturbations and the conjugate momenta is reduced to solving essentially an eigenvalue problem.

The canonical momenta for the Lagrangian \eqref{eq:multi_lagrangian} are
\begin{align}
\pi_{\mathcal{R}} &= { \partial L \over \partial \dot{\mathcal{R}}} =  2 a^3 \left(  \epsilon  \dot{\mathcal{R}} - \omega \sqrt{2\epsilon} q_{\rm n}  \right) \, , 
\\
\pi_{i} &= { \partial L \over \partial \dot{q}^i} =  a^3 \left( \dot{q}_i -  z_{ij} q^j \right) \, ,
\end{align}
where the indices $i$ and $j$ count the $\mathcal{N}-1$ isocurvature perturbations orthogonal to $\mathcal{R}$, and after performing the Legendre transformation the Hamiltonian is found as
\begin{equation}
\mathcal{H} =  {1 \over 2}  {\pi^i \pi_i \over a^3}  +  z_{ij} \pi^{i} q^j  + a {1 \over 2 } \nabla q^i  \nabla q_i  +  {1 \over 2} a^3  M_{ij} q^i q^j  + {3 \over 2} a^3 \omega^2 q_{\rm n}^2 + 2 {\omega \over \sqrt{2\epsilon}} \pi_{\mathcal{R}} q_{\rm n}  + {1 \over 4 }  {\pi_{\mathcal{R}}^2  \over\epsilon  a^3} +  {\epsilon  \over  a^2 }  \nabla \mathcal{R}  \nabla \mathcal{R}  \, .
\end{equation}
Note that the canonical momenta are related to the covariant time derivatives of the canonically normalized fields in the following manner:
\begin{align}
\pi_{\mathcal{R}} &= a^3  \sqrt{2\epsilon} \left( \uD_t q_{\rm t} - \frac{1}{2}H\eta q_{\rm t} - \omega  q_{\rm n} \right) \, , 
\\ 
\pi_i &= a^3 \left( \uD_t q_i -\omega q_{\rm n} \delta_{i{\rm n}} \right) \, .
\end{align}
The Hamilton's equations 
\begin{equation}
\dot{q}^a  = {\partial \mathcal{H}  \over \partial \pi_a} 
\quad \text{and} \quad 
\dot{\pi}_a =  - {\partial \mathcal{H}  \over \partial q^a} 
\end{equation}
are explicitly
\begin{align}
\dot{\mathcal{R}}  & =
{1 \over 2} {\pi_\mathcal{R}   \over \epsilon a^3}  + \sqrt{\frac{2}{\epsilon}} \omega q_{\rm n} \, ,  
\\ 
\dot{\pi}_\mathcal{R} & =   
2 \epsilon a \nabla^2 \mathcal{R}   \, , 
\\
\dot{q}_i & = 
{\pi_i  \over a^3}  +  z_{ij} q^j \, , 
\\ 
\dot{\pi}_i & = 
- 2 {\omega \over \sqrt{2\epsilon}} \pi_{\mathcal{R}}\delta_{i{\rm n}} +z_{ij} \pi^{j}  +  a^3 \left(   {1 \over a^2 }  \nabla^2 q_i    - M_{ij} q^j - 3 \omega^2 \delta_{i{\rm n}} q_{\rm n}  \right) \, .
\end{align}
%
%
%
%
%
%
%
%
%
%
To investigate the behaviour of perturbations and find their approximate solutions, it is more convenient to switch to the $e$-folding number $\ud N = H \ud t$ after redefining the canonical momenta as 
\begin{equation}
P_\calR \equiv \frac{\pi_\calR}{2\epsilon a^3H}
\quad \text{and} \quad
P_i \equiv \frac{\pi_i}{a^3H} \, ,
\end{equation}
and moving to Fourier space. Then we find the following Hamilton's equations:
\begin{align} \label{eq:dR}
\mathcal{R}'  & = P_\mathcal{R}  + \sqrt{\frac{2}{\epsilon}} \Omega q_{\rm n} \, , 
\\  
\label{eq:canonical_momentum}
P_\mathcal{R}' & = - (3 - \epsilon + \eta) P_\mathcal{R}   -\kappa^2 \mathcal{R}  \, , 
\\ 
q_i' & = P_i +  Z_{ij} q^j \, , 
\\ 
\label{eq:canonical_momentum_normal}
P_i' & = -  (3 - \epsilon) P_i   - 2  \sqrt{2\epsilon} \Omega P_{\mathcal{R}} \delta_{i{\rm n}} +Z_{ij} P^{j}   
- \kappa^2 q_i    - \mathcal{M}_{ij} q^j - 3 \Omega^2 \delta_{i{\rm n}} q_{\rm n} \, ,
\end{align}
with 
\begin{equation} 
\label{eq:define_kappa}
\Omega \equiv {\omega \over H}  \quad \text{and} \quad \kappa \equiv  {k \over aH} \, .
\end{equation}
The formulation in terms of the fields and their ``momenta''\footnote{Even though we will keep referring to the variables $P_i$ as momenta, these are not the canonical momenta of the problem.} nicely illustrates the superhorizon sourcing of the curvature perturbation due to a non-vanishing normal perturbation \cite{Gordon:2000hv}: Excluding runaway solutions $\mathcal{R}\propto a^2$ which would render the perturbation theory invalid, \eqref{eq:canonical_momentum} implies that for $\kappa\rightarrow0$ the momentum $P_\mathcal{R}$  is exponentially decaying. Therefore, $\mathcal{R}$ is sourced by the normal perturbation outside the horizon, whereas $q_{\rm n}$ is not. 
Moreover, for a sudden change of the bending parameters the previous set of equations has no singularities and no special junction conditions are needed at the discontinuities (as was done in e.g.~\cite{Palma:2020ejf,Fumagalli:2020nvq,Boutivas:2022qtl,Aragam:2023adu}) apart from continuity of the fields and their momenta.

\section{Analytic solutions for constant turns} 
\label{sec:analytical_sol}

To gain further physical insights, in this section we consider a simple two-field system that undergoes a sharp or broad turn. We can find a set of analytic solutions of perturbations, and in particular can identify how the positive eigenvalues enhance the perturbations.
Note that while we use Hamilton's equations to study the behaviour of the two perturbations, we have checked that our results for a single turn are consistent with~\cite{Palma:2020ejf,Fumagalli:2020nvq} for the sharp turn case once the Hubble friction is ignored. For the broad turn case we revisit some of the assumptions made in \cite{Bjorkmo:2019qno} and clarify the role of the Hubble friction terms in the derivation of the analytic solutions.



\subsection{Sharp turns} 

\label{subsec:analytical}

For two fields, the $4\times 4$ system of the fields and their conjugate momenta is the following:
\begin{align} 
\label{eq:2field-eq1}
\mathcal{R}'  & = P_\mathcal{R}  + 2\Omega  \mathcal{S} \, , 
\\
\label{eq:2field-eq2} 
P_\mathcal{R}' & = - (3 - \epsilon + \eta) P_\mathcal{R}  - \kappa^2 \mathcal{R}  \, , 
\\
\label{eq:2field-eq3}
\mathcal{S}' & = P_{\mathcal{S}}  \, , 
\\ 
\label{eq:2field-eq4}
P_{\mathcal{S}}' & =  - 2   \Omega P_{\mathcal{R}} - (3 - \epsilon + \eta) P_{\mathcal{S}} - \left[\kappa^2 + m_{\rm eff} + {1 \over 2} \eta \left( 3 - \epsilon + {1 \over 2}\eta \right) + {1 \over 2} \eta' \right] \mathcal{S} \, ,
\end{align}
%
where we used the entropic perturbation $\mathcal{S} \equiv q_{\rm n}/\sqrt{2\epsilon}$ and 
\begin{equation}
m_{\rm eff} \equiv \mathcal{M}_{nn} + 3\Omega^2\, .
\end{equation}
Now, we assume that the turn is finite, i.e.~the turn occurs within a finite duration of time and that $|\eta| \ll 1$. 
Since the duration of the turn is finite, the two perturbations are independent before the turn so we assign the Bunch-Davies initial conditions to both of them. After the turn, the two perturbations are decoupled again and are given as a superposition of Bessel functions of the first and second type:
\begin{align} 
\label{eq:after_turn_two_fields}
\mathcal{R} & = {\mathcal{R}|_{\rm te} \over \kappa_{\rm te}} \left[  \kappa \cos \left( \kappa - \kappa_{\rm te}\right)  - \sin \left( \kappa - \kappa_{\rm te} \right) \right] + {P_{\mathcal{R}}|_{\rm te}  \over \kappa_{\rm te}^3}  \left[ \left( \kappa - \kappa_{\rm te}\right) \cos \left( \kappa - \kappa_{\rm te} \right) -  (1 + \kappa  \kappa_{\rm te}) \sin \left( \kappa - \kappa_{\rm te} \right) \right] \, , 
\\
\mathcal{S}& = {\pi \kappa^{3/2}\over 2 \kappa_{\rm te}^{3/2}} \left\{ \left[ \mathcal{S}|_{\rm te} \left(\nu - {3 \over 2} \right) + P_{\mathcal{S}}|_{\rm te} \kappa_{\rm te}  \right] J_{\nu}(\kappa_{\rm te}) -  \mathcal{S}|_{\rm te} \kappa_{\rm te} J_{\nu-1}(\kappa_{\rm te}) \right\} Y_{\nu}(\kappa) 
\nonumber\\  \label{eq:after_turn_two_fields2}
& \quad
+  {\pi \kappa^{3/2}\over 2 \kappa_{\rm te}^{3/2}}  \left\{ \left[ \mathcal{S}|_{\rm te} \left(\nu - {3 \over 2} \right) + P_{\mathcal{S}}   |_{\rm te} \kappa_{\rm te}  \right] Y_{\nu}(\kappa_{\rm te}) -  \mathcal{S}|_{\rm te} \kappa_{\rm te} Y_{\nu-1}(\kappa_{\rm te}) \right\} J_{\nu}(\kappa) 
\, ,
\end{align}
where 
\begin{equation}
\nu \equiv \sqrt{{9 \over 4} - \mathcal{M}_{\rm nn}} \, ,
\end{equation}
and the subscript ``te'' denotes the values at the end of the turn. Setting $\kappa=0$ in the above relations gives the asymptotic values for the two perturbations which depend on the values of the two perturbations and their momenta at the end of the turn. To simplify the analysis we have ignored the slow-roll parameters.

During the turn the two perturbations become coupled and one has to solve the $4\times4$ non-autonomous system of equations, which we can write in matrix form $X' = J X$ with $X\equiv (\mathcal{R},q_{\rm n},P_\mathcal{R},P_{\rm n})^T$ and  $J$ the $4 \times 4$ time-dependent stability matrix. We model the turn with the step function $\Theta$:
\begin{equation}
\Omega = {\delta \theta \over \delta N} \left[ \Theta\left(N-N_{\rm ti} \right) -  \Theta\left(N-N_{\rm te} \right)  \right] \, ,
\end{equation}
where $\delta \theta \equiv \int^{N_{\rm te}}_{N_{\rm ti}} \Omega \, \ud N$ is the total angle swept during the turn  and $\delta N \equiv N_{\rm te} - N_{\rm ti}$ is the duration of the turn. The general solution at the end of the turn is given by the time-ordered exponential, which can be represented as a Dyson series:
\begin{equation} 
\label{eq:dysonseries}
X(N_{\rm te})  =  \left[I + \int^{N_{\rm te}}_{N_{\rm ti}}  \ud N_1 J(N_1) + {1 \over 2} \int^{N_{\rm te}}_{N_{\rm ti}} \int^{N_{\rm te}}_{N_{\rm ti}} \ud N_2 \ud N_1 J(N_1)  J(N_2)  + \cdots  \right] X_0 \, .
\end{equation}
Since $J$ has no oscillating part, when the duration of the turn is small we can approximate every integral with a rectangle as
\begin{equation}
I + \delta N J(N_*) + {1 \over 2} (\delta N)^2 J(N_*)  J(N_*)  + \cdots = \exp \left[ J(N_*) \delta N \right] \, ,
\end{equation}
where $N_*$ can in principle be any value lying in the interval $\left[ N_{\rm ti},N_{\rm te} \right]$ (usually the middle value yields better numerical results). The solution is then found by diagonalizing $J(N_*)$, with $J(N_*)=U^{-1}J_{\rm D} U$, and expressing it as a linear combination of the eigenvectors $\boldsymbol{e}_i$ of $U$ multiplied by exponentials:
\begin{equation}
X = \sum_i c_i  \exp (\lambda_i N) \, \boldsymbol{e}_i 
\, ,
\end{equation}
%
where $c_i$ are found from the initial conditions, the eigenvectors are
\begin{equation}
\boldsymbol{e}_i  = 
\begin{pmatrix}
1 
\vspace{0.4em}\\
- \dfrac{\kappa^2}{\bar{\lambda }_i   } 
\vspace{0.4em}\\ 
\dfrac{\kappa^2 + \lambda_i   \bar{\lambda }_i   }{ 2 \Omega  \bar{\lambda }_i  } 
\vspace{0.4em}\\ 
\lambda_i   \dfrac{\kappa^2 + \lambda_i   \bar{\lambda }_i   }{ 2 \Omega  \bar{\lambda }_i  } 
\end{pmatrix}
\, ,
\end{equation}
%
where there is no summation in $i$ and $\lambda$'s are the diagonal elements of $J_{\rm D}$, explicitly given by
\begin{align} \label{eq:eigenvalues_4}
\lambda_{1,2,3,4} & = - {1\over 2} \left[ 3-\epsilon \pm \sqrt{(3-\epsilon)^2 - 4 \kappa^2- 2 m_{\rm eff} \pm 2 \sqrt{ m_{\rm eff}^2 + 16 \kappa^2  \Omega^2 } } \right] \, , 
\\ 
\lambda_1 + \lambda_2 = \lambda_3 + \lambda_4 &= - (3 - \epsilon) \, , 
\\
\bar{ \lambda}_i & \equiv  (3 - \epsilon) + \lambda_i \, . 
\end{align}
%
It is worth noticing that if we neglect the Hubble friction terms, which is usually taken when one considers the evolution of perturbations on subhorizon scales, the corresponding solutions are the same as the above without the factor $3-\epsilon$. We will return to this point later.

%
After the turn, the two perturbations decouple so that $\mathcal{R}$ freezes on the superhorizon scales. This value is given by $\mathcal{R}|_{\rm te}$ and $P_\mathcal{R}|_{\rm te}$. Since the system consists of linear differential equations, these values in turn depend linearly on the values of the four variables at the beginning of the turn, which we denote by the subscript ``ti''. We can split $\mathcal{R}|_{\rm te}$ and $P_\mathcal{R}|_{\rm te}$ into two parts, one that we denote by the superscript $(\mathcal{R})$ being dependent on the initial curvature pair $\left(\mathcal{R}|_{\rm ti} , P_{\mathcal{R}}|_{\rm ti}\right)$, and the other denoted by $(\mathcal{S})$ being determined by the initial isocurvature pair $\left( \mathcal{S}|_{\rm ti} , P_{\mathcal{S}}|_{\rm ti}  \right)$:
\begin{align}
\mathcal{R}|_{\rm te} 
& = 
\mathcal{R}^{(\mathcal{R})}|_{\rm te} +  \mathcal{R}^{(\mathcal{S})}|_{\rm te}
=
\mathcal{R}^{(\mathcal{R})}|_{\rm te} \left( \mathcal{R}|_{\rm ti} , P_{\mathcal{R}}|_{\rm ti}  \right)    
+  \mathcal{R}^{(\mathcal{S})}|_{\rm te} \left( \mathcal{S}|_{\rm ti} , P_{\mathcal{S}}|_{\rm ti}  \right) \, , 
\\
P_{\mathcal{R}}|_{\rm te} 
& = 
P_{\mathcal{R}}^{(\mathcal{R})}|_{\rm te} + P_{\mathcal{R}}^{(\mathcal{S})}|_{\rm te}
=
P_{\mathcal{R}}^{(\mathcal{R})}|_{\rm te} \left( \mathcal{R}|_{\rm ti} , P_{\mathcal{R}}|_{\rm ti}  \right)    
+ P_{\mathcal{R}}^{(\mathcal{S})}|_{\rm te} \left( \mathcal{S}|_{\rm ti} , P_{\mathcal{S}}|_{\rm ti}  \right) \, ,
\end{align}
%
with the following explicit expressions:
\begin{align}
\mathcal{R}^{(\mathcal{R})}|_{\rm te}  
& \equiv  {\left( \kappa^2 - \lambda_3 \lambda_{4} \right) \left[ \lambda_2  e^{ \lambda_1 \delta N} \left( \lambda_1 P_{\mathcal{R}}|_{\rm ti}   - \kappa^2 \mathcal{R}|_{\rm ti} \right) -  \lambda_1 e^{ \lambda_2 \delta N} \left( \lambda_2 P_{\mathcal{R}}|_{\rm ti}   - \kappa^2 \mathcal{R}|_{\rm ti} \right) \right] \over \kappa^2 (\lambda_2 - \lambda_{4})(\lambda_1 - \lambda_{4})(  \lambda_1 - \lambda_2 )} 
\nonumber\\ 
& \quad 
+ {\left( \kappa^2 - \lambda_1 \lambda_2 \right)  \left[ \lambda_{4}  e^{ \lambda_3 \delta N} \left( \lambda_3 P_{\mathcal{R}}|_{\rm ti}   - \kappa^2 \mathcal{R}|_{\rm ti} \right) -  \lambda_3 e^{ \lambda_{4} \delta N} \left( \lambda_{4} P_{\mathcal{R}}|_{\rm ti}   -  \kappa^2 \mathcal{R}|_{\rm ti} \right) \right] \over \kappa^2 (\lambda_2 - \lambda_{4})(\lambda_1 - \lambda_{4})(  \lambda_{4} - \lambda_3 )} 
\, ,
\\
\mathcal{R}^{(\mathcal{S})}|_{\rm te} 
& \equiv 
{2 \Omega  \left[ \lambda_2  e^{ \lambda_1 \delta N} \left( P_{\mathcal{S}}|_{\rm ti} - \lambda_2 \mathcal{S}|_{\rm ti}  \right) -  \lambda_1 e^{ \lambda_2 \delta N} \left( P_{\mathcal{S}}|_{\rm ti} - \lambda_1 \mathcal{S}|_{\rm ti}  \right)  \right] \over  (\lambda_2 - \lambda_{4}) (\lambda_1 - \lambda_{4}) (  \lambda_1 - \lambda_2 )} 
\nonumber\\ 
& \quad 
+ { 2 \Omega  \left[ \lambda_4  e^{ \lambda_3 \delta N} \left( P_{\mathcal{S}}|_{\rm ti} - \lambda_4 \mathcal{S}|_{\rm ti} \right) -  \lambda_3 e^{ \lambda_4 \delta N} \left( P_{\mathcal{S}}|_{\rm ti} - \lambda_3 \mathcal{S}|_{\rm ti} \right)  \right] \over (\lambda_2 - \lambda_{4}) (\lambda_1 - \lambda_{4}) (  \lambda_{4} - \lambda_3 )} 
\, ,
\\
P_{\mathcal{R}}^{(\mathcal{R})}|_{\rm te}  
& \equiv
{\left( \kappa^2 - \lambda_3 \lambda_{4} \right)  \left[ e^{ \lambda_1 \delta N} \left( \lambda_1 P_{\mathcal{R}}|_{\rm ti}  - \kappa^2 \mathcal{R}|_{\rm ti}\right) -  e^{ \lambda_2 \delta N} \left( \lambda_2 P_{\mathcal{R}}|_{\rm ti}  - \kappa^2 \mathcal{R}|_{\rm ti} \right) \right] \over (\lambda_2 - \lambda_{4})(\lambda_1 - \lambda_{4})(  \lambda_1 - \lambda_2 )} 
\nonumber\\ 
& \quad 
+ { \left( \kappa^2 - \lambda_1 \lambda_2 \right)  \left[  e^{ \lambda_3 \delta N} \left( \lambda_3 P_{\mathcal{R}}|_{\rm ti}   - \kappa^2 \mathcal{R}|_{\rm ti} \right) -  e^{ \lambda_{4} \delta N} \left( \lambda_{4} P_{\mathcal{R}}|_{\rm ti}   - \kappa^2 \mathcal{R}|_{\rm ti} \right) \right] \over  (\lambda_2 - \lambda_{4})(\lambda_1 - \lambda_{4})(  \lambda_{4} - \lambda_3 )} 
\, ,
\\
P_{\mathcal{R}}^{(\mathcal{S})}|_{\rm te}
& \equiv 
{ 2 \Omega  \kappa^2 \left[  e^{ \lambda_1 \delta N} \left( P_{\mathcal{S}}|_{\rm ti} - \lambda_2 \mathcal{S}|_{\rm ti} \right) -  e^{ \lambda_2 \delta N} \left( P_{\mathcal{S}}|_{\rm ti} - \lambda_1 \mathcal{S}|_{\rm ti}  \right)  \right] \over (\lambda_2 - \lambda_{4}) (\lambda_1 - \lambda_{4}) (  \lambda_1 - \lambda_2 )} 
\nonumber\\ 
& \quad 
+ {2 \Omega \kappa^2  \left[  e^{ \lambda_3 \delta N} \left( P_{\mathcal{S}}|_{\rm ti} - \lambda_4 \mathcal{S}|_{\rm ti}  \right) -   e^{ \lambda_4 \delta N} \left( P_{\mathcal{S}}|_{\rm ti} - \lambda_3 \mathcal{S}|_{\rm ti}  \right)  \right] \over  (\lambda_2 - \lambda_{4}) (\lambda_1 - \lambda_{4}) (  \lambda_{4} - \lambda_3 )} 
\, .
\end{align}
Here, for $\kappa$ in principle we can use any value between $\kappa_{\rm te}$ and $\kappa_{\rm ti}$, but numerically we find that the value corresponding to $\delta N/2$ is more accurate [corresponding to the better finite difference approximation scheme of $\int  \ud N' J(N')$]. This split becomes very useful when quantizing the two perturbations.

\begin{figure}[t]
\centering
\includegraphics[width=0.7\textwidth]{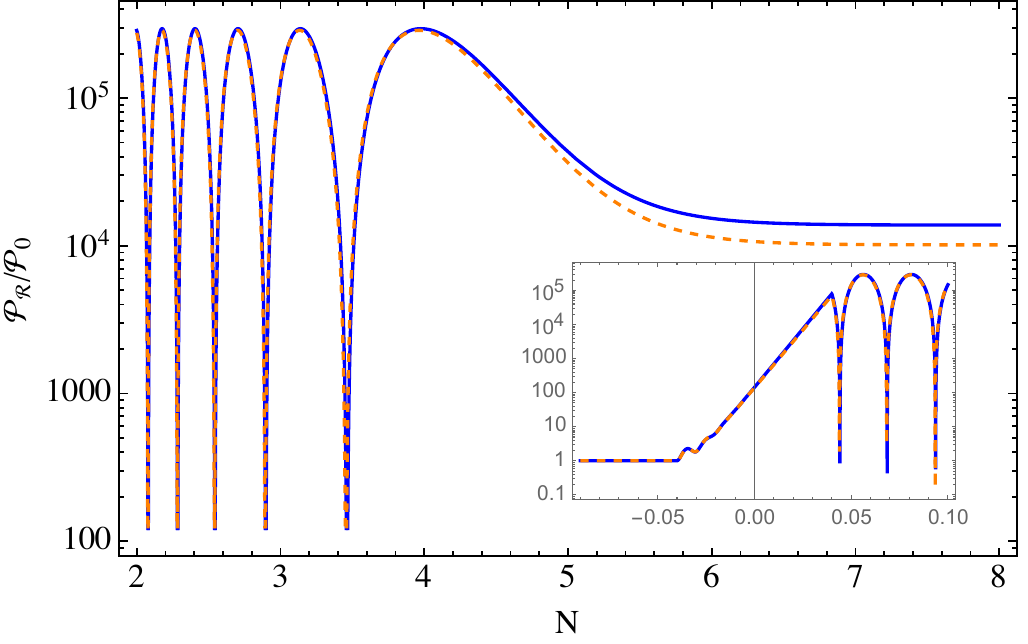}
\caption{The evolution of the curvature perturbation (normalized by the single-field expression $\mathcal{P}_{0}$) for a sharp turn (taking the form of a top-hat function) of total swapped angle $4\pi$ and duration $\delta N=0.1$ $e$-folds. The effective mass has been chosen as $m_{\rm eff} = 2 \Omega^2 + 5 $ and the turn is centred around $N=0$. The two curves correspond to the numerical solution (solid) and the approximate analytic one (dashed) given by the series of formulae presented in section~\ref{subsec:analytical}. The negligible difference after the horizon-crossing point ($N \sim 5$) is due to neglecting the slow-roll parameters in the analytic approximation.}
\label{fig:power_spectrum_one_turn_single_k}
\end{figure}

Having found the analytic solutions, we now proceed to compute the power spectrum at the end of inflation. For this goal, we promote the fields to operators and decompose them in terms of the creation and annihilation operators, and solve for the vacuum state mode functions. Then, the power spectrum is obtained by the absolute value squared of the mode function evaluated at the end of inflation. For the current case, the sensible moment to set up the operators is well before the horizon crossing where we can treat both $\mathcal{R}$ and $\mathcal{S}$ quantum mechanically. Thus, the two independent degrees of freedom at early stage are $\mathcal{R}|_\text{ti}$ and $\mathcal{S}|_\text{ti}$ that can be written in terms of the Bunch-Davies mode function solutions. Consequently, the two asymptotic values for the curvature perturbation \eqref{eq:after_turn_two_fields} for $\kappa \to 0$ can be divided into two parts with one being dependent only on $\mathcal{R}^{(\mathcal{R})}|_\text{te}$ and $P_\mathcal{R}^{(\mathcal{R})}|_\text{te}$, and the other only on $\mathcal{R}^{(\mathcal{S})}|_\text{te}$ and $P_\mathcal{R}^{(\mathcal{R})}|_\text{te}$:
%
%
%
\begin{align}
\mathcal{R}^{(\mathcal{R})}(\kappa\to0)  
&= 
{\mathcal{R}^{(\mathcal{R})} |_{\rm te} \over \kappa_{\rm te}}  \sin \left(  \kappa_{\rm te} \right) + {P_{\mathcal{R}}^{(\mathcal{R})}|_{\rm te}  \over \kappa_{\rm te}^3}  \left[  \sin \left( \kappa_{\rm te} \right)  - \kappa_{\rm te} \cos \left(  \kappa_{\rm te} \right) \right] \, ,
\\
\mathcal{R}^{(\mathcal{S})}(\kappa\to0)  
&= 
{\mathcal{R}^{(\mathcal{S})}|_{\rm te} \over \kappa_{\rm te}}  \sin \left(  \kappa_{\rm te} \right) + {P_{\mathcal{R}}^{( \mathcal{S})}|_{\rm te}  \over \kappa_{\rm te}^3}  \left[  \sin \left( \kappa_{\rm te} \right)  - \kappa_{\rm te} \cos \left(  \kappa_{\rm te} \right) \right] 
\, ,
\end{align}
from which we find the power spectrum as
\begin{equation}
\mathcal{P}_{\mathcal{R}} 
=
{ k^3 \over 2 \pi^2} \left[ \left| \mathcal{R}^{(\mathcal{R})}(\kappa\to0) \right|^2 
+  \left| \mathcal{R}^{(\mathcal{S})}(\kappa\to0) \right|^2 \right] \, .
\end{equation}
In figure~\ref{fig:power_spectrum_one_turn_single_k} we verify the accuracy of our analytic estimate for the power spectrum by plotting the amplification (relative to the single-field result) of the power spectrum for a given mode. For the specific choice of parameters we observe an amplification of almost 4 orders of magnitude, with the analytic result being perfectly consistent with full numerical one.

\subsection{Broad turns}
\label{subsec:broadturns}

When the turn is sufficiently long in duration, the situation is drastically different. Looking at the formal expression \eqref{eq:dysonseries}, it becomes clear that in principle one has to use the full Dyson series to obtain a reliable solution,\footnote{The only exception is the very restricted class of time-dependent matrices $J$ with constant eigenvectors which allows for the simple solution of the form $\exp\left( \int \ud N J \right)$.} though in most cases some perturbative scheme is assumed and the Dyson series is truncated at a given order. In this subsection we will show how this is done in the case of constant and large turns.

The first (approximate) analytic solution in the case of constant turns was presented in~\cite{Bjorkmo:2019qno} under certain assumptions. First, it relied on the effective field theory description of~\cite{Garcia-Saenz:2018vqf} which predicted a power spectrum at the end of inflation given as $\mathcal{P} = \mathcal{P}_0 e^{2c}$, with $ \mathcal{P}_0 $ being the single-field power spectrum and $c$ an unknown constant that is derived from the full two-field theory. Second, the Hubble friction terms were assumed negligible on subhorizon scales, an assumption that was also present in predecessor works on the effective field theory for constant turns, e.g.~\cite{Achucarro:2012yr}. Lastly, by pointing out the stability matrix of the coupled $\mathcal{R}$-$\mathcal{S}$ differential system (without the friction terms) acquires a positive eigenvalue at a given time before horizon crossing, the unknown constant $c$ is identified as $\int \ud N \lambda_4$, where $\lambda_4$ corresponds to the eigenvalue that becomes positive. Even though the solution was obtained as an approximation it matched numerics to a great accuracy.

However, examining the role of the Hubble friction terms in more detail leads to an apparent contradiction. If the Hubble friction terms are subdominant then on similar grounds one might expect that including those terms into the definition of the fourth eigenvalue would yield more accurate results. Surprisingly though, integrating the more accurate version of the fourth eigenvalue [which we have listed in~\eqref{eq:eigenvalues_4}] severely underestimates the overall amplification by an amount that is beyond the expected accuracy of the method. This happens because the eigenvalue including the friction terms has the form $- ( 3- \epsilon) + \sqrt{(3- \epsilon)^2 + A}$, in contrast to the one without the friction terms $\sqrt{A}$ and thus falls off faster as $A$ decreases, i.e.~as the mode approaches the horizon, leading to significantly less amplification. This also has the implication that for small $\kappa$ the dominant $\kappa$-dependent part of $\log \mathcal{P}$ takes the form $|c_{\rm s}^2| \kappa^2$ for the former case, with $c_{\rm s}^2$ being the sound speed in the effective description, as opposed to $|c_{\rm s}| \kappa$ for the latter case which is also the correct result. Therefore, to derive the correct formula one \emph{has} to omit the Hubble friction terms, which seems self-contradicting.

The role of the friction terms is important even for subhorizon scales as they control the amplitude of oscillations. These terms cannot be neglected when solving the equations for an extended period of time. For instance, ignoring Hubble friction in the equation for the single-field curvature perturbation yields the wrong result as it predicts $|\mathcal{R}| \rightarrow 0$ for $N\rightarrow -\infty$ instead of the correct value $|\mathcal{R}| \sim e^{-N}$. Even though a case could be made that these terms would yield subleading corrections for large turns, in reality for this to hold one should consider extreme turns of the order $\mathcal{O}(300)$, well outside the validity of perturbation theory. In what follows we will derive an approximate solution for $\mathcal{R}$ as well as for $\mathcal{S}$, which was not derived in the previous references, without resorting to an effective field theory approach. This will be done by converting the equations for perturbations into perturbative expansion in terms of $\kappa^{-1}$ and find a consistent solution for the zeroth-order part, essentially applying the adiabatic theorem \cite{1950JPSJ....5..435K} adjusted to linear differential equations. Moreover, we clarify the role of the Hubble friction terms which, when taken into account, amount for the correct amplitude at horizon crossing.

For the moment, 
we restrict ourselves to the familiar single-field equation for the curvature perturbation written up to leading slow-roll corrections as
\begin{equation}
 \mathcal{R}''+ 3  \mathcal{R}' + x^2  \mathcal{R}= 0 \, ,
\end{equation}
where $x$ is defined in~\eqref{eq:xdefinition} and we are interested in solving it in the subhorizon regime, i.e.~$x \gg 1$. To this end, we perform the change of variables from $N$ to $x$ to obtain
\begin{equation}
 \mathcal{R}_{,xx} - {2 \over x}  \mathcal{R}_{,x} +  \mathcal{R}= 0 \, ,
\end{equation}
which is the equation in terms of the rescaled conformal time. This equation can be written as a first-order system with a perturbative expansion in terms of $x^{-1}\ll 1$, 
\begin{equation}
\label{eq:Xx-eq}
X_{,x} = J_{(0)} X + J_{(1)} X \, ,
\end{equation}
where 
\begin{equation}
 X \equiv 
 \begin{pmatrix}
  \mathcal{R} \\
  p_{\mathcal{R}}  
 \end{pmatrix}
 \, , \qquad
  J_{(0)} \equiv 
 \begin{pmatrix}
 0 & 1\\
 -1 & 0
 \end{pmatrix}
 \, , \qquad
  J_{(1)} \equiv 
 \begin{pmatrix}
 0 & 0 \\
 0 & \dfrac{2}{x}
 \end{pmatrix} \, .
\end{equation}
The momentum with $x$ as the time variable, $p_{\mathcal{R}}$, is related to $P_{\mathcal{R}}$, i.e.~the momentum in terms of the $e$-folding time, via the chain rule $P_{\mathcal{R}} = -x \, p_{\mathcal{R}}$. The next step is to diagonalize the zeroth-order stability matrix as $J_{\rm D} \equiv U^{-1} J_{(0)} U$ and rewrite \eqref{eq:Xx-eq} in terms of a new variable $X \equiv UY$:
\begin{equation} 
\label{eq:ydiff}
Y_{,x} = \underbrace{J_{\rm D} Y}_{\text{diagonal 0\textsuperscript{th} order}}  +   \underbrace{\left(U^{-1} J_{(1)} U Y -  U^{-1} U_{,x}  \right) Y}_\text{1\textsuperscript{st} order} \, .
\end{equation}
The last term is differentiated by $x$ and is hence first-order in terms of $x^{-1}$. By using this transformation we can reduce the problem to a zeroth-order diagonal part plus a perturbation term, and our goal is to consistently solve the zeroth-order part only. This, however, does not mean we can discard entirely 
this perturbation term. 
Since the matrix $U$ is not uniquely defined, ignoring the first-order part in the previous equation and finding $Y$ to zeroth order will make $X$ ambiguous as it depends on the specific choice of $U$. 
To completely remove this ambiguity, we note that for $U$ we have the freedom to multiply each of its columns by an arbitrary function, which amounts to rescaling the eigenvectors of $J_{(0)}$. Thus we can choose $U$ in such a way that the subdominant term in~\eqref{eq:ydiff} has only non-diagonal elements. This choice can be shown to be 
\beq 
\tilde{U } = U \exp \left(\int \ud x\, P_{\rm D} \right) \, ,
\eeq
where $P_{\rm D}$ denotes the diagonal elements of the first-order part in~\eqref{eq:ydiff}, i.e.~we split it as follows:
\beq
U^{-1} J_{(1)} U Y -  U^{-1} U_{,x}   = P_{\rm D} + P_{\rm od} \, .
\eeq
Based on that observation we rewrite~\eqref{eq:ydiff} as a sum of a diagonal and an off-diagonal part:
\beq
\label{eq:ydiff2}
Y_{,x} = D_{\rm Y} Y  +  P_{\rm od} Y \, ,
\eeq
with $ D_{\rm Y} \equiv J_{\rm D} + P_{\rm D} $, which yields the leading-order solution for the fundamental matrix of our original problem as\footnote{In a sense, to derive the solution in the so called ``adiabatic picture'' we apply the interaction picture not on the original system $X' =J\, X$ but on the ``rotated system'' described by $Y= U^{-1} \, X$. Then we break the new stability matrix in two parts, the first being $J_{\rm D} + P_{\rm D} $ which we can solve analytically (since it consists of diagonal terms), and the second being the interaction part consisting of the matrix with entries the off-diagonal elements of $U^{-1} J_{(1)} U Y -  U^{-1} U_{,x}$ (see also \cite{PhysRevA.45.3329,Blanes:2008xlr}).}
\beq \label{eq:berry}
F_{\rm X} = U \exp \left[ \int \ud N D_{\rm Y}   \right] \, .
\eeq
Notice that $\int \ud N P_{\rm D}$ is related to the Berry's phase \cite{Berry:1984jv}.

For our previous example we find the eigenvectors of $J_{(0)}$ as $e_1 = (- i ,1)^T$, $e_2 = ( i ,1)^T$ and the diagonal higher-order terms of elements of \eqref{eq:ydiff} as $\text{diag}(1/x,1/x)$. Therefore, by redefining the matrix $U$ as
\begin{equation}
\tilde{U} = \begin{pmatrix}
- i x & i x \\
x & x
\end{pmatrix}
\, ,
\end{equation}
we find the fundamental matrix (i.e.~the matrix of linearly independent solutions) of the lowest-order solution as 
\begin{equation}
F_{\rm Y} = \text{diag} \left(e^{i  x} ,  e^{ - i x} \right) \, ,
\end{equation}
and subsequently the fundamental matrix of our original problem via $F_{\rm X} = \tilde{U} F_{\rm Y}$. Therefore, we find the general solution for the curvature perturbation and its momentum as
\begin{equation} 
\label{eq:approxR}
\begin{aligned}
\mathcal{R} &= c_1 x e^{ix} + c_2 x e^{-ix} \, , \\
P_\mathcal{R} &= -x \, p_\mathcal{R} = -  c_1 i \, x^2 e^{ix} + c_2 i \, x^2 e^{-ix} \, ,\
\end{aligned}
\end{equation}
which are accurate up to $\mathcal{O}(1)$ corrections. It should also be clear that the Hubble friction terms provide the correct linear dependence on $x$, even though they are suppressed in the $x^{-1}$ expansion. This happens because they contribute $\log x$ corrections and thus cannot be consistently ignored in the final result even to leading order (see e.g.~\cite{Gong:2001he} for a derivation in a similar spirit). The result obtained with the previous method coincides with the WKB solution when considering an asymptotic expansion $ \mathcal{R}  = \exp \left( x \mathcal{R}_{(0)} + \mathcal{R}_{(1)} + \cdots \right)$ and keeping only the first two terms.  
Although for the single-field case the standard WKB method is simpler, for more fields substituting the above asymptotic expressions would result into a coupled time-dependent system considerably more involved than what we obtain with the method described so far.

Moving to the two-field case, we observe that the $4\times4$ system \eqref{eq:2field-eq1}-\eqref{eq:2field-eq4}
 includes three scales, the Hubble scale $x$, the turn rate $\Omega$ and the mass parameter $m_{\rm eff}$. Again, we are interested in the subhorizon regime, as on superhorizon scales one can apply the transfer functions formalism to study the evolution of perturbations (see e.g.~\cite{Bassett:2005xm,Peterson:2010np,Peterson:2011yt,Greenwood:2012aj,Kaiser:2012ak}). Therefore, we first write the system in the $x$-parameterization:
\begin{align} 
 \mathcal{R}_{,x} & =  p_\mathcal{R}  -  {2 \Omega \over x} \mathcal{S} \, , 
\\
p_{\mathcal{R},x}  & = {2 \over x}   p_\mathcal{R}  - \mathcal{R}  \, , 
\\
\mathcal{S}_{,x} & = p_{\mathcal{S}}  \, , 
\\ 
p_{\mathcal{S},x}  & =   { 2  \Omega \over x} p_{\mathcal{R}} +  {2 \over x} p_{\mathcal{S}} - \left(1 + {m_{\rm eff}  \over x^2} \right) \mathcal{S} \, .
\end{align}
The previous set of equations describe a perturbative expansion in inverse powers of $x$ which remains valid until $x$ becomes of order $\text{max}(\sqrt{m_{\rm eff} },2\Omega,1)$, where we also assume $\Omega \gg 1$. Note that during the first phase the two perturbations decay as almost massless fields. Depending on the relative magnitude of $2 \Omega$ to $ \sqrt{m_{\rm eff}} $ there are two subsequent types of behaviour. The first case includes models with $ m_{\rm eff} \geq 4\Omega^2$ which have been studied in early works of effective field theory descriptions, e.g.~\cite{Cremonini:2010ua,Achucarro:2012yr}. For these models the amplification of the power spectrum as a function of the turn rate scales at most as $\mathcal{P}_{\mathcal{R} } \sim \sqrt{\Omega}$ and hence they are not appropriate candidates for PBH formation (see also Appendix \ref{app:masslarge}). For this reason for the remainder of this section we focus on the latter case, namely $ m_{\rm eff} < 4\Omega^2$.

To properly account for the subsequent regime we will use a new variable $\theta$, defined from $\ud \theta \equiv -2\Omega/x \ud x = 2 \Omega \ud N$,\footnote{$\theta$ has the interpretation of the angle swept in the field space during evolution.} for which the system becomes
\begin{align} \label{eq:dRtheta}
 \mathcal{R}_{,\theta} & =  \tilde{p}_\mathcal{R}  + \mathcal{S} \, , 
\\
\tilde{p}_{\mathcal{R},\theta}  & = - {3 \over 2 \Omega} \tilde{p}_\mathcal{R}  - \left({x \over 2 \Omega}\right)^2 \mathcal{R}  \, , 
\\
\mathcal{S}_{,\theta} & = p_{\mathcal{S}}  \, , 
\\  \label{eq:dpstheta}
p_{\mathcal{S},\theta}  & =   - \tilde{p}_{\mathcal{R}}   -  {3 \over 2 \Omega}  p_{\mathcal{S}} - \left[ \left({x \over 2 \Omega}\right)^2 + {m_{\rm eff}  \over 4 \Omega^2} \right] \mathcal{S} \, ,
\end{align}
where $\tilde{p}_{\mathcal{R}} \equiv P_{\mathcal{R}}/(2\Omega)$. 
Similarly to the single-field case we split the full matrix to zeroth- and first-order parts: 
\begin{equation} \label{eq:J_split_0}
 J_{(0)} \equiv 
 \begin{pmatrix}
 0 & 1 & 1& 0\\
 -\left( \dfrac{x}{2\Omega} \right)^2 &0 &0 &0 \\
 0 & 0 & 0& 1 \\
 0 & -1 &  -\left[\left( \dfrac{x}{2\Omega}\right)^2 + \dfrac{m_{\rm eff}}{4\Omega^2} \right] & 0
 \end{pmatrix}
 \quad \text{and} \quad
  J_{(1)} \equiv 
 \begin{pmatrix}
 0 & 0 & 0& 0\\
0 & - \dfrac{3}{2\Omega} &0 &0 \\
 0 & 0 & 0& 0 \\
 0 & 0 &  0 & - \dfrac{3}{2\Omega} 
 \end{pmatrix}\, .
\end{equation}
The four eigenvalues of $J_{(0)}$ are $\{\pm \lambda_{\rm im} , \pm \lambda_{\rm p} \}$ with
\begin{align} 
\label{eq:lambda_im}
\lambda_{\rm im} &\equiv {i \over 2\Omega \sqrt{2} } \sqrt{ 2 x^2 +   m_{\rm eff} +   \sqrt{ m_{\rm eff}^2 + 16 x^2  \Omega^2 } } \, , 
\\ 
\label{eq:lambda_p}
\lambda_{\rm p} & \equiv   {1\over 2\Omega \sqrt{2} } \sqrt{ - 2 x^2-  m_{\rm eff} +  \sqrt{ m_{\rm eff}^2 + 16 x^2  \Omega^2 } } \, ,
\end{align}
where the subscript ``im'' reflects the fact that this eigenvalue is always imaginary, whereas $\lambda_{\rm p}$ becomes positive for $x^2 < 4\Omega^2 - m_{\rm eff} $.  Following the same steps as before, we find the general solution for $\mathcal{R}$ and $\mathcal{S}$ as 
\begin{align} 
\label{eq:curvature_WKB}
\mathcal{R} &= x  f_+(x)  \left(c_1 e^{ - \int  \lambda_{\rm im}    }  + c_2 e^{ \int  \lambda_{\rm im}  }  \right) +  x f_{-}(x)  \left(c_3 e^{- \int  \lambda_{\rm p}  }  + c_4 e^{ \int  \lambda_{\rm p} }  \right) \, , 
\\ 
\label{eq:isocurvature_WKB}
\mathcal{S} &=   x  f_+ (x) s_{\rm im}   \left( c_1 e^{ - \int  \lambda_{\rm im}  } - c_2 e^{ \int  \lambda_{\rm im}   }  \right) +   x f_{-} (x) s_{\rm p}  (x) \left( - c_3 e^{ - \int  \lambda_{\rm p}  }  + c_4 e^{ \int  \lambda_{\rm p}  }  \right) \, ,
\end{align}
where  $f_{\pm}(x)$ and $s_{{\rm im}({\rm p})}(x)$ are given by
\begin{align}
f_{\pm} (x) &\equiv \left[ 1- {4 \Omega^2  \pm m_{\rm eff} \over \sqrt{m_{\rm eff}^2 +16 x^2 \Omega^2}} -  { 4 \Omega^2 (2x^2 \mp m_{\rm eff}) \over m_{\rm eff}^2 +16 x^2 \Omega^2}  \right]^{1/4} \, , 
\\ \label{eq:sp}
s_{{\rm im} ({\rm p})}(x) & \equiv  - {  8 \Omega^2 \lambda_{{\rm im} ({\rm p})}^2 + 2x^2 \over 8 \Omega^2 \lambda_{{\rm im} ({\rm p})} }\, .
\end{align}
The previous solution has been derived under the assumption of negligible off-diagonal terms in~\eqref{eq:ydiff} for the rotated vector $Y$. Thus, we need to check whether these terms remain small compared to one. 

First, we focus on the massless case $m_{\rm eff} =0$. The off-diagonal elements $P_{\rm od}$ remain $\mathcal{O}(\Omega^{-1})$ until horizon crossing, hence we can trust the expansion up to that order. However, $P_{\rm od}$ diverge as $x\rightarrow 2\Omega$ which is the point where the eigenvalues $\pm \lambda_{\rm p}$ switch from imaginary to positive/negative, indicating that $J_{(0)}$ becomes non-diagonalizable at that point (the analogue of a turning point in the WKB approximation). This further implies that there will be a small region around $x=2\Omega$ where the approximate solution has to be found with a different method. Nevertheless, it suffices to consider the two solutions as Bunch-Davies modes and match the two expressions at a later time. 
Moving to masses satisfying $0<m_{\rm eff} < 4\Omega^2$ the range of validity of the above approximate solution is more limited. The off-diagonal elements diverge as $x\rightarrow \sqrt{4\Omega^2-m_{\rm eff}}$, which again is the point where $\lambda_{\rm p}$ crosses zero. In addition, as $x\rightarrow \sqrt{2\Omega}$ the off-diagonal terms become  $\mathcal{O}(\Omega^{-1/2})$ and eventually scale as $\sqrt{4\Omega^2}/\sqrt{4\Omega^2-m_{\rm eff} }$ when $x \rightarrow 1$. This means that the approximate solution \eqref{eq:isocurvature_WKB} is strictly valid for $x\in \left(\sqrt{2\Omega},x_* \right)$ with $x_*<\sqrt{4\Omega^2-m_{\rm eff}}$. For the scales $x\in(1,\sqrt{2\Omega})$ a more accurate version would require a different split of zeroth and first-order terms or another rotation $X = U Y$, where now $U$ diagonalizes a different matrix. However, simply using the previous formulae we find a deviation of at most a factor of 10 closer to the borderline limit. Specifically for $m_{\rm eff} =4\Omega^2$ the analytical solution completely fails (Appendix \ref{app:masslarge} shows how to obtain the correct solution in this case).

\begin{figure}[t]
\centering
\includegraphics[width=0.49\textwidth]{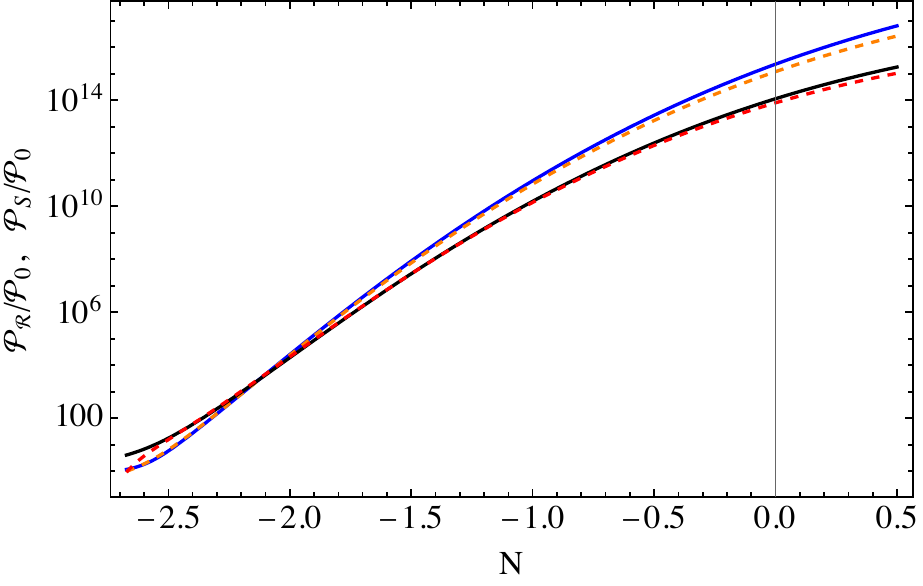}
\includegraphics[width=0.49\textwidth]{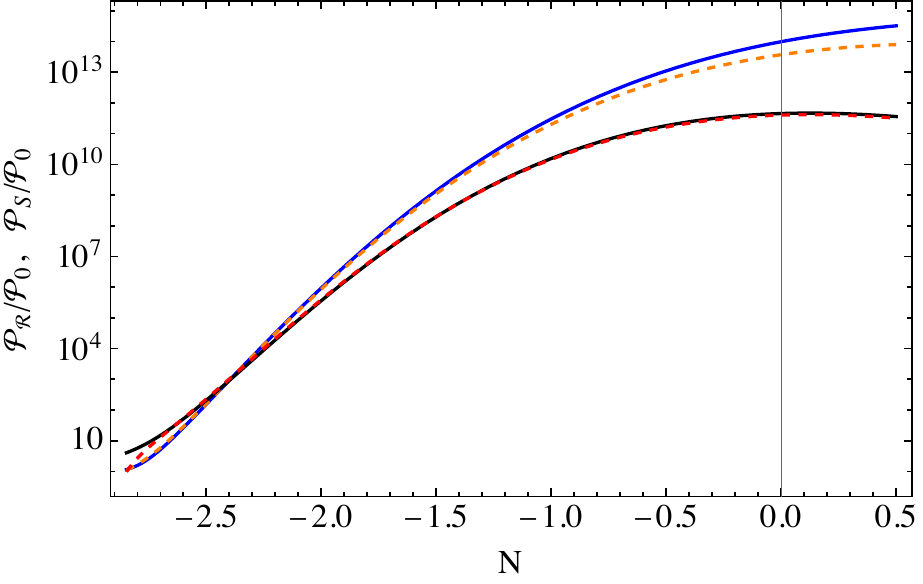}
\caption{Numerical (solid) versus analytic approximation (dashed) for the curvature (blue solid-orange dashed) and isocurvature (black solid-red dashed) power spectra (normalized by the single-field expression $\mathcal{P}_{0}$). Even though we have considered moderate values for the turn rate the approximation remains accurate up to horizon crossing ($N=0$). We have checked that larger values of the turn rate improve the accuracy of the analytic approximation. 
We consider in the left panel the massless case $m_{\rm eff} = 0$ with $\Omega=8$, and in the right panel $m_{\rm eff} = 2\Omega^2$ with $\Omega=11$.}
\label{fig:massless}
\end{figure}

Regarding the constants $c_i$ these can be found by matching \eqref{eq:curvature_WKB} and  \eqref{eq:isocurvature_WKB} with the values of the two perturbations before the point $x_*$, e.g.~the Bunch-Davies solutions. For our purposes it is sufficient to treat them as $\mathcal{O}(1)$ numbers and hence at horizon crossing we quote the final result for the curvature perturbation:
\begin{equation}
\left.\mathcal{P}_\mathcal{R}  \right|_{x=1}  \approx \mathcal{P}_0 |f_-|^2 
\left. \exp \left[ 2 \int_{\sqrt{4\Omega^2 - m_{\rm eff}}}^1 \ud x \left(-{2\Omega \over x} \right) \lambda_{\rm p} \right] \right|_{x=1} \, ,
\end{equation}
with $\mathcal{P}_0\approx x^2$ for $x\approx 1$. Note that extending the upper integration limit in the above integral to $x\rightarrow 0$  yields the exponential amplification factor $\pi \Omega \left(2 - \sqrt{m_{\rm eff}} / \Omega\right)$ that was mentioned in \cite{Bjorkmo:2019qno}. We have plotted the approximate solution in figure \ref{fig:massless} for two choices of the mass parameter and by considering $c_3=c_4=1$ and $c_1=c_2=0$.

\section{Criteria for the growth of perturbations}  
\label{sec:hamilton}

Having explored the two-field case we move on to investigate the growth of perturbations for an arbitrary number of fields. 
%
The set of the $2\mathcal{N}$ Hamilton's equations can be written in the matrix form $X' = J X$ with $X\equiv (\mathcal{R},q_i,P_\mathcal{R},P_i)^T$ and  $J$ being a $2\mathcal{N}\times2\mathcal{N}$ time-dependent matrix. The behaviour of perturbations is qualitatively different on sub/super-horizon scales so we will study the two regimes separately.

%
%
%

\subsection{Subhorizon evolution}

For $\mathcal{N}$ fields, we first redefine the time variable from $N$ to $x$ and obtain the following system in terms of fields and momenta:
\begin{align}
\mathcal{R}_{,x} & = p_\mathcal{R}  - \sqrt{\frac{2}{\epsilon}}  {\Omega \over x}q_{\rm n} \, , 
\\  
p_{\mathcal{R},x} & = {2 \over x} p_\mathcal{R}  -  \mathcal{R}  \, , 
\\ 
q_{i,x} & = p_i -  {Z_{ij} \over x} q^j \, , 
\\ 
p_{i,x} & =  {2 \over x} p_i  + {2  \sqrt{2\epsilon} \Omega \over x} p_{\mathcal{R}} \delta_{i{\rm n}} - {Z_{ij}\over x} p^{j}   
-  q_i    - {\mathcal{M}_{ij} \over x^2} q^j -  {3\Omega^2 \over x^2} \delta_{i{\rm n}} q_{\rm n} \, ,
\end{align}
where the indices $i$ and $j$ stand for the $\mathcal{N}-1$ isocurvature perturbations orthogonal to $\mathcal{R}$, but $a$ and $b$ denote all the $\mathcal{N}$ perturbations, and $P_a = -x \, p_a$. Moreover, we have ignored slow-roll corrections, i.e.~$\epsilon,|\eta|\ll1$. Similar to section~\ref{subsec:broadturns} we split $J$ in two parts, $J=J_{(0)}+J_{(1)}$. 
Considering large bending parameters $\Omega,T,\cdots$ only the friction terms $2p_a/x$  fall in $J_{(1)}$ and thus we neglect the friction terms and focus on $J_{(0)}$. Next, we introduce two enlarged, $\mathcal{N}\times\mathcal{N}$ matrices $\tilde{\mathcal{M}}_{ab}$ and $\tilde{Z}_{ab}$ defined as
\begin{equation}
\tilde{\mathcal{M}}_{ab} \equiv 
\begin{pmatrix}
0 & 0 \\ 
0 & \mathcal{M}_{ij} +  3 \Omega^2 \delta_{i \rm n} \delta_{j \rm n}  \\
\end{pmatrix} 
\quad \text{and} \quad 
\tilde{Z}_{ab} \equiv 
\begin{pmatrix}
0 & 0 \\ 
0 & Z_{ij} \\
\end{pmatrix} \, , 
\end{equation}
where the first row and first column correspond to the curvature perturbation $\mathcal{R}$. This renders $J$ in the following block-diagonal form:
\begin{equation}
J_{(0)} = \begin{pmatrix}
- \dfrac{1}{x} \tilde{Z}_{ab} - \dfrac{1}{x} \sqrt{\dfrac{2}{\epsilon}} \Omega \delta_{a\mathcal{R}}\delta_{b \rm n}  &  I_{ab} 
\vspace{0.3em}\\
- I_{ab}  - \dfrac{1}{x^2} \tilde{\mathcal{M}}_{ab} &  - \dfrac{1}{x} \tilde{Z}_{ab} + \dfrac{2}{x} \sqrt{2 \epsilon} \Omega \delta_{a \rm n} \delta_{b\mathcal{R}}
\end{pmatrix} 
\equiv 
\begin{pmatrix}
A & B\\
C & D
\end{pmatrix} \, .
\end{equation}
%
The characteristic polynomial of $J_{(0)}$ is then generally given by
\begin{equation}
\label{eq:eigenvalues_n_fields}
\lambda^{2\mathcal{N}}  - \text{Tr}(J_{(0)}) \lambda^{2\mathcal{N}  -1 } + \dots + \det(J_{(0)}) = 0 \, ,
\end{equation}
and so, by the Routh-Hurwitz theorem,\footnote{The Routh-Hurwitz theorem states that whenever a coefficient of this characteristic polynomial becomes negative, then at least one eigenvalue acquires positive real part (see e.g.~\cite{2008arXiv0802.1805B}).} a sufficient condition for the existence of a growing mode is $\det(J_{(0)})<0$.\footnote{Note that for $\mathcal{N}>2$ this is only a sufficient condition. One can imagine the case of two negative eigenvalues and the rest $\mathcal{N}-2$ positive giving overall a positive determinant.} 
The determinant of $J_{(0)}$ can be given as the determinant of a matrix constructed by multiplying the four blocks using the following formula:
\begin{equation}
\text{det} (J_{(0)}) = \text{det} (A)  \det \left(D - C A^{-1} B \right) =   \det \left(DA - C  \right)  \, , 
\end{equation}
since in our case $B$ is the identity matrix, and the matrix we have to calculate is given explicitly as
\begin{equation}
\label{eq:dac}
DA - C =  I_{ab} + {1\over x^2} \left(\tilde{\mathcal{M}}_{ab} - 4 \Omega^2 \delta_{a\rm n} \delta_{b \rm n} + \tilde{Z}_{ak} \tilde{Z}_{kb} \right)   \, .
\end{equation}
We observe that for two fields this matrix becomes a number equal to 
$1 + \left(m_{\rm eff} -4 \Omega^2\right)/x^2$.

The determinant can further be simplified by noticing that the previous matrix is block diagonal with 1 as its first entry and zeros along the rest elements of its first row and column:
\begin{equation}
\det (J_{(0)}) =  \det \left[ I_{ij} + {1\over x^2}\left(\mathcal{M}_{ij} - 4 \Omega^2 \delta_{i\rm n} \delta_{j \rm n} + \tilde{Z}_{ik} Z_{kj} \right) \right] 
\equiv
\det \left(  I + {1\over x^2}  \mathcal{M}_{\rm sub}\right)
\, ,
\end{equation}
where $\mathcal{M}_{\rm sub}$ generalizes the effective mass of the isocurvature perturbation on subhorizon scales, and 
the indices $i$ and $j$ only count for these $\mathcal{N}-1$ isocurvature perturbations.
With the expression of the determinant at our disposal we are ready to predict when the perturbations will be amplified. Using the eigenvalues $\mu_i$ of the matrix $\mathcal{M}_{\rm sub}$, 
%
we can factorize the determinant in the following way:
\begin{equation}
\det(J_{(0)})  = \prod_i \left(  1+ { \mu_i  \over x^2}\right) \, .
\end{equation}
If some of these eigenvalues are complex then they should come in pairs and the above factorization becomes
\begin{equation}
\det(J_{(0)}) =  \prod_m \left(  1 + {\mu_m \over x^2}  \right)  \prod_j \left( 1 + {2\over x^2} \Re(\mu_j)+ {|\mu_j|^2\over x^4} \right) \, ,
\end{equation}
where now $m< \mathcal{N}-1$ and $j < (\mathcal{N}-1)/2 $ stand for the numbers of real and complex eigenvalues respectively, with $m+j=\mathcal{N}-1 $.
For negative determinant at least one of the eigenvalues of $ \mathcal{M}_{\rm sub} $ should have negative real part.

We also mention that the eigenvalue equation \eqref{eq:eigenvalues_n_fields} can be obtained from the determinant of the following matrix:
\begin{equation}
\det(J_{\rm eig,sub} ) \equiv \det \left(\lambda^2 I - (A+D)\lambda +DA - C  \right) =0 \, .
\end{equation}
%
This expression can further be simplified using the Laplacian expansion of the determinant in terms of its minors. Expanding along its first row the determinant is expressed as
\begin{equation} 
\label{eq:jeig_determinant}
\begin{aligned}
\det(J_{\rm eig,sub} ) =& \left( \lambda^2 + 1 \right) \det \left[ (\lambda^2 + 1) I + {2 \over x} Z \lambda+ {1 \over x^2}\mathcal{M}_{\rm sub}  \right] 
\\
& + {4 \Omega^2 \lambda^2 \over x^2} \det \left(  \left[ (\lambda^2 + 1) I + {2 \over x} Z \lambda+ {1 \over x^2}\mathcal{M}_{\rm sub} \right]_{\rm nn} \right)  =0 \, .
\end{aligned}
\end{equation}
In the case of two fields this equation reproduces the four eigenvalues mentioned in \eqref{eq:lambda_im} and \eqref{eq:lambda_p} of section~\ref{subsec:broadturns} after setting the cofactor $[\cdots]_{\rm nn}$ to 1.

\subsection{Superhorizon evolution} 
\label{subsec:superhorizon}

On superhorizon scales the system of \eqref{eq:dR}-\eqref{eq:canonical_momentum_normal} is already in perturbative form and so we can isolate the $\kappa$-dependent part as
\begin{equation}
X ' =  J_{\rm sup (0)} X + J_{\rm sup (1)} X \, .
\end{equation}
Diagonalizing $ J_{\rm sup (0)} $ as  $J_{\rm sup(0)} = S^{-1 }J_{\rm eig (0)} S$, with $J_{\rm eig(0)} $ a diagonal matrix and $Y \equiv S X$, the above equation can equivalently be written as
\begin{equation}
Y ' =  J_{\rm eig (0)} Y + S J_{\rm sup (1)} S^{-1} Y \, .
\end{equation}
The second term is proportional to $\kappa^2$ and remains finite when integrated in the interval $(0,\kappa_i)$ and so (by Levinson's theorem \cite{levinson}) stability can be inferred by focusing on the eigenvalues of the $\kappa$-independent part. Thus, we can set $\kappa=0$ in the original system and focus on the eigenvalues of $J_{\rm sup (0)} $. Due to the structure of $J_{\rm sup (0)} $ the evolution equations for $\mathcal{R}$ and $P_{\mathcal{R}}$ decouple from the rest  equations so we can focus entirely on the orthogonal subspace. The stability matrix describing the orthogonal perturbations is explicitly
\begin{equation}
J_{\rm orth,sup}  = \begin{pmatrix}
Z_{ij}  & I_{ij} 
\vspace{0.3em}\\
-  \mathcal{M}_{ij} - 3 \Omega^2 \delta_{i \rm n} \delta_{j \rm n}&  - (3-\epsilon ) I_{ij} +  Z_{ij} 
\end{pmatrix} \, ,
\end{equation}
%
and the eigenvalue equation becomes 
%
\begin{equation}
\det(J_{\rm eig,sup} )= \det \left(\lambda^2 I + \left( (3 -\epsilon) I - 2 Z  \right) \lambda + \mathcal{M}_{\rm sup} \right) =0 \, ,
\end{equation}
where we have defined for later convenience the superhorizon mass matrix for the isocurvature perturbations:
\begin{equation}
\left( \mathcal{M}_{\rm sup} \right)_{ij} \equiv \mathcal{M}_{ij} + 3 \Omega^2 \delta_{i \rm n} \delta_{j \rm n} + Z_{ik}Z_{kj} - (3-\epsilon ) Z_{ij} \, .
\end{equation}
Similarly, we can express the eigenvalue equation in terms of the minor determinants as follows:
\begin{equation}
\begin{aligned}
\left[ \lambda^2  + (3-\epsilon)\lambda + \mathcal{M}_{\rm nn} + 3 \Omega^2 -T^2 \right] \det \left( \left[J_{\rm eig,sup}\right]_{\rm nn}  \right) &\\
- \left[ \mathcal{M}_{\rm nb}^2 - \left( 2 \lambda + 3-\epsilon \right)^2T^2 \right] \det \left( \left[ J_{\rm eig,sup}\right]_{\rm nn,bb} \right) &= 0 
\end{aligned} \, .
\end{equation}
To achieve stability on superhorizon scales the eigenvalues of this equation should have negative real part.

For two fields $J_{\rm orth,sup}$ is just a number and the eigenvalue equation becomes simply $\lambda^2  + (3 -\epsilon) \lambda +m_{\rm eff}=0$ with the following solutions:
\begin{equation}
\lambda_{1,2} =  - {1\over 2} \left[ 3-\epsilon \pm \sqrt{(3-\epsilon)^2 - 4m_{\rm eff}} \right] \, ,
\end{equation}
which have negative real part if $m_{\rm eff}>0$. For three fields we find four eigenvalues which we can express in terms of the trace and determinant of $\mathcal{M}_{\rm sup} $ as
\begin{equation}
\label{eq:lambda_i_sh}
\lambda_{i} =  - {1\over 2} \left\{ 3-\epsilon \pm \sqrt{(3-\epsilon)^2 - 2 \text{Tr}(\mathcal{M}_{\rm sup} ) -8 T^2 \pm 2 \sqrt{ \left[ \text{Tr}(\mathcal{M}_{\rm sup} ) + 4 T^2 \right]^2 - 4 \det(\mathcal{M}_{\rm sup} )} } \right\} \, .
\end{equation}
Demanding $\Re(\lambda_{i}) < 0$ we find the conditions for stable perturbations listed in \cite{Christodoulidis:2022vww}. Similar to the turn rate, the presence of the torsion provides a stabilizing effect on the orthogonal perturbations.


\section{Models with isometries} 
\label{sec:isometries}

Models with isometries have attracted a lot of attention in the literature. Not only they yield simpler models, but they are appropriate candidates of rapid-turn or rapid-twist behaviour~\cite{Christodoulidis:2022vww}. Moreover, the isometry structure linearly relates the nb-component of the mass matrix with the torsion through $\mathcal{M}_{\rm nb}=(3-\epsilon)T$. In this section we will demonstrate that the bb-component satisfies $\mathcal{M}_{\rm bb}=T^2$, which implies that for three fields the mass matrix has only one free parameter that does not automatically depend on the kinematical quantities, and then we will focus on the case of massless isocuvature perturbations.

\subsection{$\mathcal{N}-1$ isometries}
\label{subsec:isometries_N}

For $\mathcal{N}-1$ isometries the metric tensor can be written in block diagonal form with no cross-couplings between the non-isometric field $\chi$ and the isometric fields. For this coordinate system the torsion vector can be found from the expression
\begin{equation} 
\label{eq:tbi}
T b^i \approx \Gamma^i_{\chi j}v^j + \Omega t^i   \, ,
\end{equation}
which squared provides us with an expression for the torsion:
\begin{equation}
T^2 + \Omega^2 = \Gamma^i_{\chi k} \Gamma^j_{\chi l} v^kv^l \, .
\end{equation}
Moreover, notice that the tangent and torsion vectors have zero components in the $\chi$ direction, and for an attractor solution the mixed potential derivatives with respect to the isometry fields should be zero:
\begin{equation}
(\log V)_{,ij} \approx 0 \, .
\end{equation}
This is a trivial statement for one field whereas for more fields this is proved by contradiction: If $(\log V)_{,ij} \neq 0$ then one can find non-trivial solutions to  
\begin{equation}
\label{eq:no_go}
{\ud \phi^i \over \ud \phi^j } =  {V^{,i} \over V^{,j} } \, ,
\end{equation}
implying some parametric relations between the fields that depend on the initial conditions $\phi^i = \phi^i(\phi^j,\phi^j_0)$ and hence to a non-unique late-time solution. Combining the previous two allows us to simplify the element $\mathcal{M}_{\rm bb}$ as
\begin{equation}
M_{\rm bb}  = V_{\rm bb}  -  R_{kijl} \dot{\phi}^i\dot{\phi}^jb^kb^l \approx - b^k b^l \Gamma_{kl}^{\chi} V_{,\chi} -  G_{kd} \left( \Gamma^{d}_{i\chi} \Gamma^{\chi}_{l j} -  \Gamma^{d}_{l \chi} \Gamma^{\chi}_{i j} \right) \dot{\phi}^i \dot{\phi}^j b^k b^l  \approx T^2 H^2\, ,
\end{equation}
where we made use of \eqref{eq:tbi} and the vanishing of the effective gradient $V_{,\chi} + \Gamma^{\chi}_{ij} \dot{\phi}^i \dot{\phi}^j \approx 0$.

Because of this form of the components of the mass matrix, the $x$-dependent eigenvalues which follow from \eqref{eq:jeig_determinant}, greatly simplify up to slow-roll corrections. 
For three fields they are given by
\begin{align} \label{eq:eigenvalue_3_isometry}
\lambda_{\pm \pm} & =  \pm {1\over \sqrt{2} x} \sqrt{ -2 x^2- \mu_{\rm eff} \pm  \sqrt{ \mu_{\rm eff}^2 + 16 x^2 \left(T^2 + \Omega^2 \right) + 4 (3-\epsilon)^2T^2} } \, , 
\\
\lambda_{5,6} &= \pm i \, ,
\end{align}
with
\begin{equation}
\mu_{\rm eff} \equiv \mathcal{M}_{\rm nn} + 3\Omega^2 + 3 T^2 \, .
\end{equation}
The  eigenvalue $\lambda_{++}$ becomes positive when 
\begin{equation}
\label{eq:isometries_amplification}
x^2 + \mathcal{M}_{\rm nn} - \Omega^2 -  T^2  - {(3-\epsilon)^2 T^2 \over x^2}< 0 \, ,
\end{equation}
which, for relatively large values of $\Omega$ and $T$, marks the onset of the amplification (see figure~\ref{fig:ultralight}). Notice that the last term in \eqref{eq:isometries_amplification} is suppressed on subhorizon scales and, hence, this three-field model is very similar to a two-field model with a turn rate given by $\Omega^2_{\rm 2f} = \Omega^2 + T^2$; this was also noticed in \cite{Aragam:2023adu} for similar choices of the mass matrix elements but with $\mathcal{M}_{\rm nb}=0$, which is the reason why the last term in \eqref{eq:eigenvalue_3_isometry} was absent. On superhorizon scales by investigating \eqref{eq:lambda_i_sh}, the condition for stability reduces to simply $\mu_{\rm eff}>0$.

\subsection{A toy model with massless isocurvature perturbations}

As we saw in section~\ref{subsec:superhorizon} on superhorizon scales the orthogonal perturbations admit exponentially decaying solutions with exponents given by \eqref{eq:lambda_i_sh}. We will call the perturbations massless when all eigenvalues become zero. For three fields, in particular, the two eigenvalues become zero only if 
\begin{equation}
\label{eq:massless-conditions}
\det(\mathcal{M}_{\rm sup} ) = 0 
\quad \text{and} \quad 
\text{Tr}(\mathcal{M}_{\rm sup} ) + 4 T^2= 0 \, .
\end{equation}
One way for this to happen is to consider field spaces with $\mathcal{N}-1$ isometries that, as we saw in the previous section, automatically satisfy
\begin{equation}
\label{eq:mass_matrix_elements}
\mathcal{M}_{\rm bb} = T^2   
\quad \text{and} \quad 
\mathcal{M}_{\rm nb} = (3 - \epsilon) T \, , 
\end{equation}
thus fulfilling the first condition of \eqref{eq:massless-conditions}. Demanding the vanishing of the second condition fixes the last component of the mass matrix to $\mu_{\rm eff}=0$ or $m_{\rm eff} = - 3T^2 $, akin to the two-field ultra-light scenario \cite{Achucarro:2016fby} where the torsion is identically zero and the effective mass of the orthogonal field is also zero, $m_{\rm eff} = 0$. Having fixed the geometric properties of the field space, we now turn our attention to the potential.

To understand the conditions that can lead to such a potential we will use the effective potential construction \cite{Christodoulidis:2019mkj}. This assumes that a number of fields follow their potential gradient while the rest fields are frozen at the minima of their effective potential. As shown in \cite{Christodoulidis:2022vww}, a non-zero torsion in three-field models is possible when considering e.g.~a diagonal metric 
\begin{equation}
G_{IJ} = \text{diag}(f^2, g^2, 1) \, ,
\end{equation}
where $f = f(\chi)$ and $g = g(\chi)$, and two evolving fields. The vanishing of the effective gradient of the orthogonal field gives
\begin{equation}
\label{eq:effective_gradient}
V_{,\chi} - {f_{,\chi} \over f^3} (w_1)^2 H^2- {g_{,\chi} \over g^3} (w_2)^2 H^2 = 0\, ,
\end{equation}
where $w_1$ and $w_2$ are the projections of the potential gradient along the two Killing vectors and are independent of $\chi$ for a sustained slow-roll inflation with isometries.  Similarly, the vanishing of the effective mass yields
\begin{equation}
\label{eq:effective_gradient_2}
V_{,\chi \chi} + 3{f_{,\chi}^2 \over f^4} (w_1)^2 H^2 + 3 {g_{,\chi}^2 \over g^4} (w_2)^2  H^2 - {f_{,\chi \chi} \over f^3} (w_1)^2 H^2- {g_{,\chi \chi} \over g^3} (w_2)^2 H^2 =0 \, ,
\end{equation}
where we have approximated
$H_{,\chi} \approx 0$ which is equivalent to the vanishing of the effective gradient. Using the relations 
\begin{align}
{f_{,\chi}^2 \over f^4} (w_1)^2 + {g_{,\chi}^2 \over g^4} (w_2)^2 & \approx T^2 + \Omega^2  \, , 
\\
{f_{,\chi \chi} \over f^3} (w_1)^2 H^2 +  {g_{,\chi \chi} \over g^3} (w_2)^2 H^2 & \approx - R_{{\rm n} IJ {\rm n}} \dot{\phi}^I \dot{\phi}^J \, ,
\end{align}
we find that \eqref{eq:effective_gradient_2} is exactly equal to $\mu_{\rm eff}  = 0$. The problem has now reduced to finding a potential that satisfies \eqref{eq:effective_gradient} and \eqref{eq:effective_gradient_2}.

The simplest way is to consider that \eqref{eq:effective_gradient} is satisfied identically, 
integrating which implies that the potential satisfies
\begin{equation}
V + {1 \over 2} {(w_1)^2 H^2  \over f^2} + {1 \over 2}  {(w_2)^2 H^2  \over g^2} = c(\phi^1,\phi^2)\, .
\end{equation}
Comparing this with the Friedmann constraint $3H^2 = G_{IJ} \dot {\phi}^I \dot {\phi}^J/2 + V$ 
allows us to make the identification on the attractor solution $c(\phi^1,\phi^2) = 3 H^2 $. Finally, trading the derivatives of the potential with those of the Hubble parameter gives
\begin{equation}
w_{I} = 2 {H_{,I} \over H} - { H^2 G^{KJ}w_{KI}w_J \over 3 - \epsilon } \, ,
\end{equation}
and discarding the last term [see the discussion around \eqref{eq:no_go}] we finally obtain the form of the potential
\begin{equation}
V = 3 H^2 - 2 G^{IJ}H_{,I} H_{,J} \, ,
\end{equation}
with $H$ being independent of $\chi$. 
Any such potential will lead to massless isocurvature perturbations on the attractor solution. More generally, the previous derivation is an example of the superpotential method in inflation (see also \cite{Achucarro:2019pux,Aragam:2019omo} for multi-field constructions).  In this method it is assumed that the velocities in the evolution of the Hubble parameter are given as
\begin{equation}
\dot{\phi}^I = -2G^{IJ}{H_{,J}} \, ,
\end{equation}
and the form of the potential follows after substituting this choice into the Friedman constraint. It can be checked by direct substitution that the mass matrix elements for this specific expression of the potential agree with the expressions \eqref{eq:mass_matrix_elements} and $\mu_{\rm eff}=0$. Finally, making a choice for the field metric and the Hubble parameter $H(\phi^1,\phi^2)$ leads to specific values of the background quantities $\epsilon$, $\Omega$ and $T$.

For two fields  on superhorizon scales one finds vanishing momenta $P_{\rm n}=P_{\mathcal{R}}=0$ and a constant isocurvature perturbation $q_{\rm n} = c$, with $c$ depending on the background quantities, which continuously feeds the curvature perturbation through the relation
\begin{equation}
\label{eq:curvature_feeding}
P_{\mathcal{R}} = \mathcal{R} ' - \sqrt{\frac{2}{\epsilon}} \Omega q_{\rm n} = 0 \, .
\end{equation}
For three fields if the torsion is zero then, in addition to the previous relations, we also find for the binormal perturbation $q_{\rm b}'$ and  $P_{\rm b}$ both vanish. This means that the binormal perturbation effectively decouples from the other two perturbations and freezes. However, for a non-zero torsion examining the orthogonal subspace we find the late time solution as
\begin{equation}
q_{\rm n}' = 0 \, , \qquad P_{\rm b} ' = 0 \, , \qquad P_{\rm b} = - T q_{\rm n} \, , \qquad P_{\rm n} = - T q_{\rm b} \, ,
\end{equation}
with $q_{\rm n}$ and $q_{\rm b}$ being non-zero, implying that the two orthogonal perturbations are unstable. Note that this does not contradict the superhorizon analysis of section~\ref{subsec:superhorizon} where it was found that the torsion in general has a stabilizing effect because here multiple eigenvalues have zero real part and any arguments based on linearization fail. Regarding the curvature perturbation the relation \eqref{eq:curvature_feeding} still holds in the presence of torsion and we obtain the additional relation
\begin{equation}
q_{\rm b} ' = - 2 T q_{\rm n} \, .
\end{equation}
Therefore, when there are two massless orthogonal perturbations the normal perturbation freezes on superhorizon scales and continuously feeds the curvature and binormal perturbations.

\begin{figure}[t]
\centering
\includegraphics[width=0.7\textwidth]{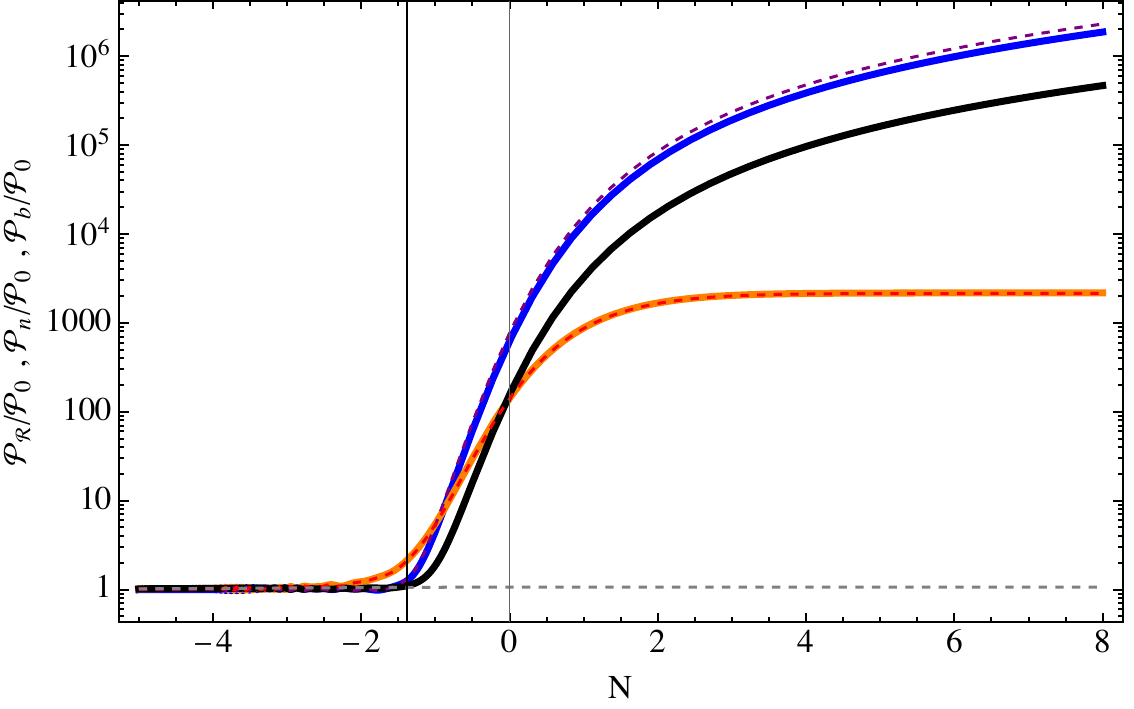}
\caption{The power spectra (normalized by the single-field spectrum ${\cal{P}}_{0}$) for a given mode that crosses the horizon at $N=0$. The turning rate and torsion are zero before $N=-3$, and become constant afterwards. The three perturbations, $\mathcal{R}$, $q_{\rm n}/\sqrt{2\epsilon}$ and $q_{\rm b}/\sqrt{2\epsilon}$, all with the Bunch-Davies initial conditions, are depicted in thin dashed purple, red and gray respectively for the model without torsion, and in thick solid blue, orange and black respectively for the model with torsion. After the horizon crossing the normal perturbation $q_{\rm n}$ freezes whereas the curvature perturbation $\mathcal{R}$ continues to increase. The behaviour of the binormal perturbation $q_{\rm b}$ depends on the value of the torsion: For the zero torsion case it is effectively decoupled from the other two and behaves as a single-field perturbation, whereas for the non-zero case it continues to increase on superhorizon scales. The vertical line at $N\approx -1.5$ corresponds to the $e$-folding time when \eqref{eq:isometries_amplification} is satisfied and the exponential amplification of perturbations commences. }
\label{fig:ultralight}
\end{figure}

In figure~\ref{fig:ultralight} we verify numerically our previous arguments regarding the behaviour of the three perturbations and their corresponding momenta (which we do not plot here) for a model with $\epsilon=0.01$ and for two sets of the bending parameters: $\Omega=\sqrt{5} \Theta(N-3)$, $T=0$ (dashed) and $\Omega=2\Theta(N-3)$, $T=\Theta(N-3)$ (solid), where $\Theta(\cdots)$ is the step function. Notice also that for both sets of parameters the exponential amplification on subhorizon scales begins when the inequality \eqref{eq:isometries_amplification} is satisfied. We have chosen the turn rate and torsion such that $\Omega^2+T^2$ is the same for both models and we observe that for the two models the curvature and normal perturbations behave almost identically (see also the discussion at the end of section~\ref{subsec:isometries_N}). However, the second model predicts sizable entropic perturbations at the end of inflation. 

\section{Conclusions}
\label{sec:conclusions}

Multiple fields are ubiquitous in models inspired by high-energy theories of the early universe. Besides their non-trivial dynamics they may display interesting phenomenologies with large deviations from the standard slow-roll predictions. In the rapid-turn regime this class of models has been shown to generate an exponential amplification of the power spectrum on small scales, while keeping almost scale invariance on the CMB scales, a mechanism that can be used for PBH formation. In this work we explored this idea further, and below we summarize our main results.

More specifically, in section~\ref{sec:multi_general} we introduced the basic ingredients of multi-field perturbations and showed that a simple change of the variable $x \equiv -k \tau$ and a field redefinition by $\tilde{v}_i =\sqrt{k} v_i$ trivially yields the scale dependence of the power spectrum when every background quantity (except for $\epsilon$ and $\eta$) is constant. The power spectrum for these models becomes identical to that of single-field models, a result which follows  simply by inspecting the equations for perturbations without resorting to an effective field theory or to the slow-turn limit. We also pointed out that the behaviour of perturbations becomes more transparent in the Hamiltonian description. Then, in section~\ref{sec:analytical_sol} we quantified the enhancement of perturbations for two-field systems that undergo constant turns of short and long duration, explaining how the two cases have to be treated differently. More specifically, for sharp turns the solution is found by exponentiating the stability matrix of perturbations, while for broad turns one has to first write the system into perturbative expansion over $1/\Omega$ and then try to solve it perturbatively. In the latter case for solutions of the form $A \exp(B)$ the phase $B$, which can be either imaginary or real, is found by solving the part of the equations without the Hubble friction terms, while the correct amplitude is found only once the Hubble friction terms are taken into account. Although we have limited our discussion on two fields, it is straightforward to generalize our results for more fields.

Subsequently, in section \ref{sec:hamilton} we investigated under what general conditions the curvature perturbation becomes exponentially enhanced on subhorizon scales, and when the perturbations remain stable on superhorizon scales, listing a series of conditions. We applied our findings on three-field models with isometries in section~\ref{sec:isometries} and showed that the amplification happens when the quantity $\mathcal{M}_{\rm nn} -\Omega^2 -T^2$ becomes negative, very analogous to the two-field case. We continued with a discussion of the ultra-light scenario with all three fields being massless on superhorizon scales and showed that when the torsion of the trajectory in the field space is non-zero then, in addition to the curvature perturbation, the binormal perturbation is also continuously sourced by the normal perturbation, the latter being constant. This partly answers the question of whether the presence of the torsion can produce distinct observational signatures in models with more than two fields.
%
We also refer the interested readers to~\cite{Aragam:2023adu} where the authors studied the amplification of the power spectrum for three-field systems due to a sudden increase of the turn rate and torsion.

\acknowledgments{It is a pleasure to thank George Kodaxis, Maria Mylova and Nikolaos Tetradis for useful discussions at the early stages of this work. We would also like to thank Gonzalo Palma and Spyros Sypsas for useful discussions during the International Molecule-Type Workshop ``Revisiting cosmological non-linearities in the era of precision surveys (YITP-T-23-03)'' at the Yukawa Institute for Theoretical Physics, Kyoto, Japan. 
We are supported in part by the National Research Foundation of Korea Grant 2019R1A2C2085023. JG also acknowledges the Korea-Japan Basic Scientific Cooperation Program supported by the National Research Foundation of Korea and the Japan Society for the Promotion of Science (2020K2A9A2A08000097). JG is further supported in part by the Ewha Womans University Research Grant of 2022 (1-2022-0606-001-1) and 2023 (1-2023-0748-001-1). 
JG is also grateful to the Asia Pacific Center for Theoretical Physics for its hospitality while this work was in progress.}

\appendix

\section{Massive field in quasi-de Sitter space} 
\label{app:massive_field}

The curvature perturbation in single-field inflation is a massless scalar field in the sense that it freezes on superhorizon scales. To find the approximate analytic solution it is better to switch to the canonical field $u=\sqrt{2\epsilon} \mathcal{R}$ which obeys the Mukhanov-Sasaki equation \cite{Sasaki:1986hm,Mukhanov:1988jd}:
\begin{equation}
\label{eq:muk_sas}
u_{,\tau\tau} + \left[k^2 - (aH)^2\left( 2  - \epsilon + {3 \over 2} \eta -  {1 \over 2} \epsilon \eta +  {1 \over 4} \eta^2 +  {1 \over 2} \eta \xi  \right) \right] u =0 \, ,
\end{equation}
with $\xi \equiv (\log \eta)'$. The term $aH$ is related to the conformal time via
\begin{equation}
 \tau - \tau_{\rm r}= \int^N_{N_{\rm r}} \ud {N}' {1 \over a H} =  \int^N_{N_{\rm r}} \ud {N}' {1 \over a_0}e^{-{N}' + \int \ud n \epsilon(n) } \, ,
\end{equation}
where the last expression follows from the definition of the slow-roll parameter $\epsilon \equiv -(\log H)'$ and for de Sitter space we follow the convention $\tau_{\rm r}\rightarrow 0$ for $N_{\rm r} \rightarrow \infty$. Integrating by parts twice yields
\begin{equation}
 \tau - \tau_{\rm r}=  - {1 + \epsilon \over a H } \Big|^N_{N_{\rm r}}+ \int \ud {N}' {\epsilon^2 + \epsilon \eta \over a H } \, , 
\end{equation}
where the second term is at least one order higher in the slow-roll parameters.\footnote{By successive integration by parts one can derive the relation of the conformal time in terms of the Hubble radius to arbitrary order in the slow-roll parameters.} To first order in the slow-roll parameters \eqref{eq:muk_sas} becomes 
\begin{equation}
u_{,\tau\tau} + \left[k^2 - {1 \over \tau^2}\left( 2  + 3\epsilon + {3 \over 2} \eta \right) \right] u =0 \, ,
\end{equation}
with solutions the H\"ankel functions of first and second type. Choosing the positive frequency solution at early times yields 
\begin{equation}
u = {1 \over 2}\sqrt{\pi} e^{i(2\nu +1) \pi/4}\sqrt{-\tau} H^{(1)}_{\nu}(\tau) \, ,
\end{equation}
with $\nu = 3/2 +\epsilon + \eta/2$. The same form of the solution holds for a massive field with mass term $\mathcal{M}_{\rm nn}$ upon replacing $\nu$ with 
\begin{equation}
\nu = \sqrt{{9 \over 4} - \mathcal{M}_{\rm nn}} \, .
\end{equation}

\section{Sub- to superhorizon evolution for models with $m_{\rm eff}\geq4\Omega^2\gg1$} 
\label{app:masslarge}

In this section we will examine how our previous method works for models when a moderate amplification on subhorizon scales is possible.

\subsection{$m_{\rm eff}>4\Omega^2$}

First, we mention models where the mass parameter is strictly larger than $4\Omega^2$, which is equivalent to a positive effective mass on subhorizon scales $m_{\rm sub}^2 = m_{\rm eff} - 4\Omega^2>0$. At scales $x\sim \sqrt{m_{\rm eff}}$ the two perturbations start deviating from the Bunch-Davies values (assuming they are initially in such states) and decrease at a faster rate than the Bunch-Davies solution. The curvature perturbation freezes on superhorizon scales but approaches its asymptotic value faster than single-field models, behaving as a single-field perturbation in the limit $m_{\rm eff} \rightarrow \infty$, see e.g.~\cite{Achucarro:2010da,Achucarro:2012sm,Achucarro:2012yr,Cremonini:2010ua}. Since these models do not yield any significant amplification, we ignore them and proceed to the borderline case.

\subsection{The borderline case $m_{\rm eff}= 4\Omega^2$}

\begin{figure}[t]
\centering
\includegraphics[width=0.7\textwidth]{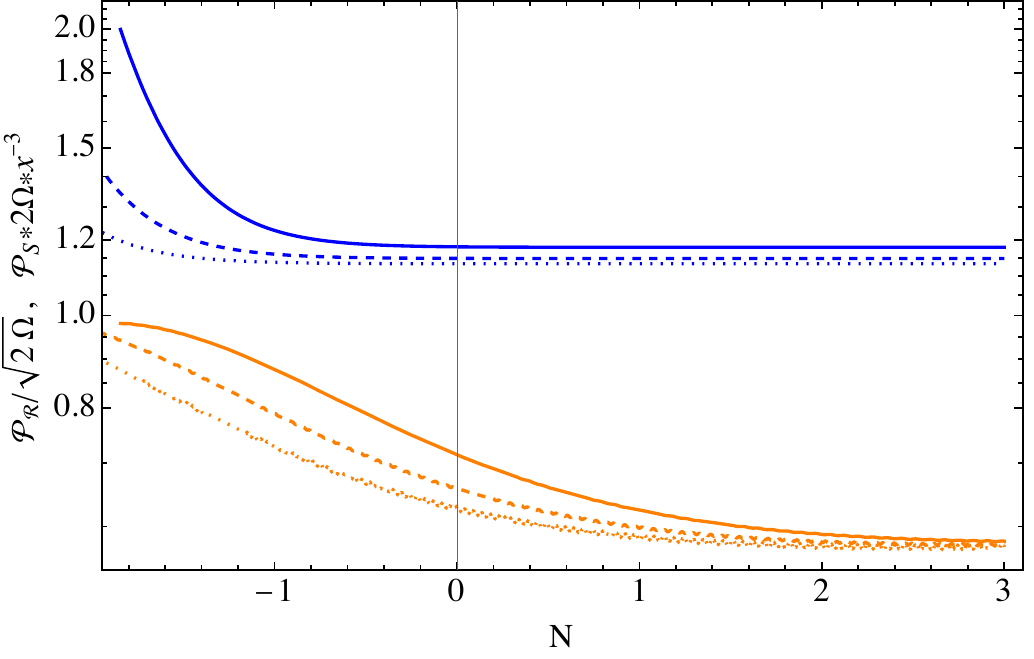}
\caption{The curvature (blue) and isocurvature (orange) power spectra normalized by their asymptotic expressions for the borderline case $m_{\rm eff}=4\Omega^2$ in the regime $x<\sqrt{2\Omega}$. Each curve is the numerical evolution for different values of the turn rate: solid, dashed and dotted curves represent $\Omega=$20, 50 and 100 respectively. The horizon crossing point is set at $N=0$.}
\label{fig:borderline}
\end{figure}

Following the same steps as in Section~\ref{subsec:broadturns} we notice a peculiarity for the borderline mass case $m_{\rm eff} = 4\Omega^2$. Splitting the stability matrix as in~\eqref{eq:J_split_0}, i.e. writing 
\beq
Y_{,\theta}  = (J_{\rm D} + P_{\rm D})Y + P_{\rm od} Y \, ,
\eeq
we find that the off-diagonal terms for the rotated system, represented by the matrix $P_{\rm od}$, become $\mathcal{O}(1)$ when $x\sim \sqrt{2\Omega}$ and then keep increasing until horizon crossing. Thus, we can not trust this approximation after this scale. Note also that choosing to diagonalize the full stability matrix instead of its zeroth-order part does not alter the previous behaviour. Using the formulae listed in Section~\ref{subsec:broadturns} we find that at $x\sim \sqrt{2\Omega}$ the four variables have almost equal amplitudes $ \sim \Omega^{1/4}$ modulo some $\mathcal{O}(1)$ constants.  Since on scales  $x\sim \sqrt{2\Omega}$ the quadratic terms $x^2/ (2\Omega)^2$ in \eqref{eq:dRtheta}-\eqref{eq:dpstheta} become $\mathcal{O}(\Omega^{-1})$, this implies they can be moved to the stability matrix of the first-order part, suggesting the following split:
\begin{equation}  
\label{eq:tildematrices}
  \tilde{J}_{(0)} \equiv 
 \begin{pmatrix}
 0 & 1 & 1& 0\\
 0 &0 &0 &0 \\
 0 & 0 & 0& 1 \\
 0 & -1 &  -1 & 0
 \end{pmatrix}
 \quad \text{and} \quad
  \tilde{J}_{(1)} \equiv 
 \begin{pmatrix}
 0 & 0 & 0& 0\\
-\left( \dfrac{x}{2\Omega} \right)^2 & - \dfrac{3}{2\Omega} &0 &0 \\
 0 & 0 & 0& 0 \\
 0 & 0 &  -\left( \dfrac{x}{2\Omega}\right)^2 & - \dfrac{3}{2\Omega} 
 \end{pmatrix}\, .
\end{equation}
By doing so we are able to solve the system consistently. In particular, we find that the curvature perturbation freezes at $x\sim\sqrt{2\Omega}$, $|\mathcal{R}|_{x <\sqrt{2\Omega}} \sim  |\tilde{p}_{\mathcal{R}} |_{x=\sqrt{2\Omega}}\sim (2\Omega)^{1/4}$ (recovering the result of \cite{Cremonini:2010ua}), whereas the isocurvature perturbation decays as $x^{3/2}/\sqrt{2 \Omega}$ (see figure \ref{fig:borderline}).

\end{document}